\documentclass[prd,showpacs,preprintnumbers,amsmath,amssymb]{revtex4}

\usepackage{graphics}
\usepackage{epsfig}
\usepackage{subfigure}
\usepackage{dcolumn}
\usepackage{bm}
\usepackage{color}

\begin{document}
\draft
\title{{\bf Dynamics of Hadronic Molecule in One-Boson Exchange Approach and Possible Heavy Flavor Molecules }}

\author{Gui-Jun Ding$^{a}$}
\author{Jia-Feng Liu$^{a}$}
\author{ Mu-Lin Yan$^{a,b}$}

\affiliation{\centerline{$^a$Department of Modern
Physics,}\centerline{University of Science and Technology of
China,Hefei, Anhui 230026, China}\centerline{$^b$ Interdisciplinary
Center for Theoretical Study,} \centerline{University of Science and
Technology of China,Hefei, Anhui 230026, China}}

\begin{abstract}
\vskip0.5cm

We extend the one pion exchange model at quark level to include the
short distance contributions coming from $\eta$, $\sigma$, $\rho$
and $\omega$ exchange. This formalism is applied to discuss the
possible molecular states of $D\bar{D}^{*}/\bar{D}D^{*}$,
$B\bar{B}^{*}/\bar{B}B^{*}$, $DD^{*}$, $BB^{*}$, the
pseudoscalar-vector systems with $C=B=1$ and $C=-B=1$ respectively.
The "$\delta$ function" term contribution and the S-D mixing effects
have been taken into account. We find the conclusions reached after
including the heavier mesons exchange are qualitatively the same as
those in the one pion exchange model. The previous suggestion that
$1^{++}$ $B\bar{B}^{*}/\bar{B}B^{*}$ molecule should exist, is
confirmed in the one boson exchange model, whereas $DD^{*}$ bound
state should not exist. The $D\bar{D}^{*}/\bar{D}D^{*}$ system can
accomodate a $1^{++}$ molecule close to the threshold, the mixing
between the molecule and the conventional charmonium has to be
considered to identify this state with X(3872). For the $BB^{*}$
system, the pseudoscalar-vector systems with $C=B=1$ and $C=-B=1$,
near threshold molecular states may exist. These bound states should
be rather narrow, isospin is violated and the $I=0$ component is
dominant. Experimental search channels for these states are
suggested.

\end{abstract}

\maketitle

\section{Introduction\label{sec:introduction}}

Since 1970s it is widely believed that Quantum Chromodynamics should
accomodate a richer spectrum than just $q\bar{q}$ and $qqq$
resonances, many possible nonconventional structures are suggested,
e.g. glueballs ($gg$, $ggg$,...), hybrid mesons ($q\bar{q}g$) and
multiquark states($qq\bar{q}\bar{q}$, $qqqq\bar{q}$, $qqqqqq$,
$qqq\bar{q}\bar{q}\bar{q}$). Unfortunately, so far there is still no
uncontroversial evidence for nonconventional states experimentally
except the hadronic molecules. The deuteron is a well-known example
of hadronic molecule, and the approximate $10^5$ known nuclear
levels are all hadronic molecule. In the past few years, many new
states have been reported, a striking feature is that some of them
are close to the thresholds of certain two hadrons, which inspires
the possible interpretation of hadronic molecule.

Hadronic molecule is an old idea, about thirty years ago the
possible hadronic molecules consisting of two charm mesons are
suggested\cite{Voloshin:1976ap}, and $\psi(4040)$ was proposed to be
a P wave $D^{*}\bar{D}^{*}$ molecule \cite{De Rujula:1976qd}. Since
in general molecule is weakly bound, the separation between the two
hadrons in the molecule should be large. We can picture the two
hadrons as interacting via a meson exchange potential
\cite{Sakurai:1960ju}. At large distance, one pion exchange is
dominant. Guided by the binding of deuteron, Tornqvist performed a
systematic study of possible deuteronlike two-meson bound states
\cite{Tornqvist:1991ks,Tornqvist:1993ng}. The role of pion exchange
in forming hadronic molecules was studied by Ericson and Karl
\cite{Ericson:1993wy}. Recently Close et al. \cite {Thomas:2008ja}
performed a pedagogic analysis of the overall sign, in addition they
included the contribution of the "$\delta$ function" term which
gives a $\delta$ function in the effective potential when no
regularization is used. In these original work, only long distance
one pion exchange has been considered, and the short distance
contributions are neglected. In Ref. \cite{swanson} Swanson assumed
that the short distance dynamics is governed by the one gluon
exchange induced constituent quark interchange mechanism, which
results in state mixing.

In the model of the nucleon-nucleon interaction, the long range part
of the nucleon-nucleon force is quantitatively accounted for by the
one $\pi$ exchange. However, the short and intermediate range
interactions are governed by more complex dynamics. Combining the
well-established one $\pi$ exchange with the exchange of heavier
bosons (e.g. scalar and vector mesons) to describe the behavior at
short distance has been proved to be a very successful approach
\cite{Nagels:1975fb,Nagels:1977ze,Machleidt:1987hj}. Physically, the
scalar and vector meson exchange describes part of multiple pion
exchange effects. For the two $\pi$ exchange, if they interact and
correlate in a P wave state, such a exchange can be modeled by
$\rho$ exchange. If the two correlated $\pi$ pair is in a S-wave
state, Durso et al. showed that one can approximate them by the
exchange of a scalar $\sigma$ meson \cite{Durso:1980vn}. Similarly,
the correlated 3 $\pi$ exchange can be approximated by the exchange
of one $\omega$ meson.

Inspired by the nucleon-nucleon interaction, we shall represent the
short distance interactions by the heavier bosons $\eta$, $\sigma$,
$\rho$ and $\omega$ exchange instead of the quark interchange. The
effective potential between two hadrons is obtained by summing over
the interactions between light quarks or antiquarks as in the
original work
\cite{Tornqvist:1991ks,Tornqvist:1993ng,Thomas:2008ja}. It is
well-known that one pion exchange between two light quarks results
in two terms: the isospin dependent spin-spin interaction and tensor
force. After taking into account the heavy bosons exchange, six
additional terms appear including the spin-isospin independent
central term, only isospin dependent term, isospin independent
spin-spin interaction and tensor force, both isospin dependent and
independent spin-orbit interactions. Consequently the situation
becomes more complex than the only one pion exchange model. In our
model, both the "$\delta$ function" term and the S-D mixing effects
would be considered, which have be shown to play an important role
in the binding
\cite{Tornqvist:1991ks,Tornqvist:1993ng,Thomas:2008ja}. In this
work, we first give a good description of the deuteron in our model,
which is an unambiguous hadronic molecule, then apply this formalism
to the heavy flavor pseudoscalar-vector (PV) systems. Thus the
predictions for the possible heavy flavor PV molecules are base on a
solid and reliable foundation. This is a greater advantage over
other approaches dealing with the dynamics of hadronic molecule,
such as one boson exchange in the effective field theory
\cite{Ding:2008gr,Liu:2008fh} and residual strong force with
pairwise interactions \cite{Wong:2003xk,Ding:2008mp} etc.

The paper is organized as follows. In section II, the formalism of
the one boson exchange model is presented, the effective potentials
from pseudoscalar, scalar and vector meson exchange are given
explicitly. In section III, we give the meson parameters involved in
our model and the boson-quark couplings which are extracted from the
boson-nucleon couplings. The formalism is applied to the deuteron in
section IV, the $D\bar{D}^{*}/\bar{D}D^{*}$ system and the molecular
interpretation of X(3872) are investigated in section V. We further
apply the one boson exchange approach to other heavy flavor PV
systems in section VI, and possible molecular states are discussed.
Section VII is our conclusions and discussions section. The
expressions for the matrix elements of the spin relevant operators
are analytically given in the Appendix.

\section{The formalism of one-boson exchange model}

The construction of one-boson exchange interaction is constrained by
the symmetry principle. To the leading order in the boson fields and
their derivative, the effective interaction Lagrangian describing
the coupling between the constituent quarks and the exchange boson
fields is as follows
\cite{Nagels:1975fb,Nagels:1977ze,Machleidt:1987hj}
\begin{eqnarray}
\nonumber \rm
{Pseudoscalar:}&&~~~~~\mathcal{L}_p=-g_{pqq}\bar{\psi}(x)i\gamma_5\psi(x)\varphi(x)\\
\nonumber \rm
{Scalar:}&&~~~~~\mathcal{L}_s=-g_{sqq}\bar{\psi}(x)\psi(x)\phi(x)\\
\label{1}\rm{Vector:}&&~~~~~\mathcal{L}_v=-g_{vqq}\bar{\psi}(x)\gamma_{\mu}\psi(x)v^{\mu}(x)-\frac{f_{vqq}}{2m_{q}}\bar{\psi}(x)\sigma_{\mu\nu}\psi(x)\partial^{\mu}v^{\nu}(x)
\end{eqnarray}
Here $m_q$ is the constituent quark mass, $\psi(x)$ is the
constituent quark Dirac spinor field, $\varphi(x)$, $\phi(x)$ and
$v^{\mu}(x)$ are the isospin-singlet pseudoscalar, scalar and vector
boson fields respectively. In this work we take $m_q\equiv
m_u=m_d\simeq313$ MeV, since we concentrate on constituent up and
down quarks. If the isovector bosons are involved, the couplings
enter in the form $\bm{\tau\cdot\varphi}$, $\bm{\tau\cdot\phi}$ and
$\bm{\tau\cdot v^{\mu}}$ respectively, where $\bm{\tau}$ is the
well-known Pauli matrices. For the pseudoscalar, another interaction
term is allowed
$\mathcal{L}'_p=\frac{f_{pqq}}{m_{p}}\bar{\psi}(x)\gamma^{\mu}\gamma_5\psi(x)\partial_{\mu}\varphi(x)$
, where $m_p$ is the exchange pseudoscalar mass, this Lagrangian has
been used by Tornqvist \cite{Tornqvist:1991ks,Tornqvist:1993ng} and
Close \cite{Thomas:2008ja}. By partial integration and using the
equation of motion, one can easily show that $\mathcal{L}_p$ and
$\mathcal{L}'_p$ are equivalent provided the coupling constants are
related by
\begin{equation}
\label{2}\frac{f_{pqq}}{m_p}=\frac{g_{pqq}}{2m _q}
\end{equation}

From the above effective interactions, the effective potential
between two quarks in momentum space can be calculated
straightforwardly following the standard procedure. To the leading
order in $\mathbf{q^2}/m^2_q$, where $\mathbf{q}$ is the momentum
transfer, the potentials are
\begin{enumerate}
\item{Pseudoscalar boson exchange}
\begin{eqnarray}
\nonumber V_p(\mathbf{q})&=&-\frac{g^2_{pqq}}{4m^2_q}\frac{(\bm{\sigma}_i\cdot\mathbf{q})(\bm{\sigma}_j\cdot\mathbf{q})}{\mathbf{q}^2+\mu^2_p}\\
\label{3}&&=-\frac{g^2_{pqq}}{12m^2_q}\left[\frac{\mathbf{q}^2}{\mathbf{q}^2+\mu^2_p}\;\bm{\sigma}_i\cdot\bm{\sigma}_j+\frac{\mathbf{q}^2S_{ij}(\hat{\mathbf{q}})}{\mathbf{q^2}+\mu^2_p}\right]
\end{eqnarray}
where
$S_{ij}(\hat{\mathbf{q}})=3(\bm{\sigma}_i\cdot\hat{\bm{q}})(\bm{\sigma}_j\cdot\hat{\bm{q}})-\bm{\sigma}_i\cdot\bm{\sigma}_j$,
we have used $\mu^2_p=m^2_p-q^2_0$ instead of $m^2_p$ to
approximately account for the recoil effect
\cite{Tornqvist:1991ks,Tornqvist:1993ng,Thomas:2008ja}.
\item{Scalar boson exchange }
\begin{eqnarray}
\label{4}V_s(\mathbf{q})=-\frac{g^2_{sqq}}{\mathbf{q}^2+\mu^2_s}\big(1+\frac{\mathbf{q}^2}{8m^2_q}\big)-\frac{g^2_{sqq}}{2m^2_q}\frac{i\mathbf{S}_{ij}\cdot(\mathbf{p}\times\mathbf{q})}{\mathbf{q}^2+\mu^2_s}
\end{eqnarray}
where
$\mathbf{S}_{ij}\equiv\frac{1}{2}(\bm{\sigma}_i+\bm{\sigma}_j)$,
$\mu^2_s=m^2_s-q^2_0$ with $m_s$ the exchange scalar meson mass, and
$\mathbf{p}$ denotes the total momentum.
\item{Vector boson exchange}
\begin{eqnarray}
\nonumber V_v(\mathbf{q})&=&\frac{g^2_{vqq}}{\mathbf{q}^2+\mu^2_v}-\frac{g^2_{vqq}+4g_{vqq}f_{vqq}}{8m^2_q}\frac{\mathbf{q}^2}{\mathbf{q}^2+\mu^2_v}+\frac{(g_{vqq}+f_{vqq})^2}{12m^2_q}\frac{\mathbf{q}^2S_{ij}(\hat{\mathbf{q}})-2\mathbf{q}^2(\bm{\sigma}_i\cdot\bm{\sigma}_j)}{\mathbf{q}^2+\mu^2_v}\\
\label{5}&&-\frac{3g^2_{vqq}+4g_{vqq}f_{vqq}}{2m^2_q}\;\frac{i\mathbf{S}_{ij}\cdot(\mathbf{p}\times\mathbf{q})}{\mathbf{q}^2+\mu^2_v}
\end{eqnarray}
where $\mu^2_v=m^2_v-q^2_0$ approximately reflects the recoil effect
with $m_v$ the exchange vector meson mass.
\end{enumerate}
The effective potential in configuration space is obtained by
Fourier transforming the momentum space potential.
\begin{equation}
\label{6}V_i(\mathbf{r})=\frac{1}{(2\pi)^3}\int
d^3\mathbf{q}\;e^{i\mathbf{q}\cdot\mathbf{r}}V_i(\mathbf{q})
\end{equation}
where $i=p$, $s$ and $v$ respectively. However, the resulting
potentials are singular, which contains delta function, so the
potentials have to be regularized. Considering the internal
structure of the involved hadrons, one usually introduces form
factor at each vertex. Here the form factor is taken as
\begin{equation}
\label{7}F(q)=\frac{\Lambda^2-m^2}{\Lambda^2-q^2}=\frac{\Lambda^2-m^2}{X^2+\mathbf{q}^2}
\end{equation}
where $\Lambda$ is the so-called regularization parameter, $m$ and
$q$ are the mass and the four momentum of the exchanged boson
respectively with $X^2=\Lambda^2-q^2_0$. The form factor suppresses
the contribution of high momentum, i.e. small distance. The presence
of such a form factor is dictated by the extended structure of the
hadrons. The parameter $\Lambda$, which governs the range of
suppression, can be directly related to the hadron size that is
approximately proportional to $1/\Lambda$. However, since the
question of hadron size is still very much open, the value of
$\Lambda$ is poorly known phenomenologically, and it is dependent on
the model and application. In the nucleon-nucleon interaction, the
$\Lambda$ in the range 0.8-1.5 GeV has been used to fit the data.
For the present application to heavy mesons, which have a much
smaller size than nucleon, we would expect a larger regularization
parameter $\Lambda$. We can straightforwardly obtain the effective
potentials between two quarks in configuration space. For
convenience, the following dimensionless functions are introduced.
\begin{eqnarray}
\nonumber H_0(\Lambda,m_{ex},\mu,r)&=&\frac{1}{\mu r}\big(e^{-\mu
r}-e^{-Xr}\big)-\frac{\Lambda^2-m^2_{ex}}{2\mu X}\,e^{-Xr}\\
\nonumber H_1(\Lambda,m_{ex},\mu,r)&=&-\frac{1}{\mu r}\big(e^{-\mu
r}-e^{-Xr}\big)+\frac{X(\Lambda^2-m^2_{ex})}{2\mu^3}\,e^{-Xr}\\
\nonumber H_2(\Lambda,m_{ex},\mu,r)&=&\big(1+\frac{1}{\mu
r}\big)\frac{1}{\mu^2r^2}e^{-\mu
r}-\big(1+\frac{1}{Xr}\big)\frac{X}{\mu}\frac{1}{\mu^2r^2}e^{-Xr}-\frac{\Lambda^2-m^2_{ex}}{2\mu^2}\frac{e^{-Xr}}{\mu
r}\\
\nonumber H_3(\Lambda,m_{ex},\mu,r)&=&\big(1+\frac{3}{\mu
r}+\frac{3}{\mu^2r^2}\big)\frac{1}{\mu r}e^{-\mu
r}-\big(1+\frac{3}{Xr}+\frac{3}{X^2r^2}\big)\frac{X^2}{\mu^2}\frac{e^{-Xr}}{\mu
r}-\frac{\Lambda^2-m^2_{ex}}{2\mu^2}\big(1+Xr\big)\frac{e^{-Xr}}{\mu r}\\
\nonumber
G_1(\Lambda,m_{ex},\tilde{\mu},r)&=&\frac{1}{\tilde{\mu}r}\;\big[\cos(\tilde{\mu
r})-e^{-Xr}\big]+\frac{X(\Lambda^2-m^2_{ex})}{2\tilde{\mu}^3}e^{-Xr}\\
\nonumber G_3(\Lambda,m_{ex},\tilde{\mu},r)&=&-\big[\cos(\tilde{\mu
r})-\frac{3\sin{(\tilde{\mu}r)}}{\tilde{u}r}-\frac{3\cos(\tilde{\mu}r)}{\tilde{\mu}^2r^2}\big]\frac{1}{\tilde{\mu}r}-\big(1+\frac{3}{Xr}+\frac{3}{X^2r^2}\big)\frac{X^2}{\tilde{\mu}^2}\frac{e^{-Xr}}{\tilde{\mu}r}\\
\label{8}&&-\frac{\Lambda^2-m^2_{ex}}{2\tilde{\mu}^2}\big(1+Xr\big)\frac{e^{-Xr}}{\tilde{\mu}r}
\end{eqnarray}
Then the effective potentials between two quarks from one-boson
exchange are
\begin{enumerate}
\item{Pseudoscalar boson exchange}
\begin{equation}
\label{9}V_p(\mathbf{r})=\left\{\begin{array}{cc}
\frac{g^2_{pqq}}{4\pi}\frac{\mu^3_p}{12m^2_q}\big[-H_1(\Lambda,m_{p},\mu_p,r)\,\bm{\sigma}_i\cdot\bm{\sigma}_j+H_3(\Lambda,m_{p},\mu_p,r)S_{ij}(\hat{\mathbf{r}})\big],&\mu^2_p>0\\
\frac{g^2_{pqq}}{4\pi}\frac{\tilde{\mu}^3_p}{12m^2_q}\big[-G_1(\Lambda,m_{p},\tilde{\mu}_p,r)\,\bm{\sigma}_i\cdot\bm{\sigma}_j+G_3(\Lambda,m_{p},\tilde{\mu}_p,r)S_{ij}(\hat{\mathbf{r}})\big],&~~\mu^2_p=-\tilde{\mu}^2_p<0
\end{array}\right.
\end{equation}
with
$S_{ij}(\hat{\mathbf{r}})=3(\bm{\sigma}_i\cdot\hat{\mathbf{r}})(\bm{\sigma}_j\cdot\hat{\mathbf{r}})-\bm{\sigma}_i\cdot\bm{\sigma}_j$
\item{Scalar boson exchange}
\begin{equation}
\label{10}V_s(\mathbf{r})=-\mu_s\frac{g^2_{sqq}}{4\pi}\left[H_0(\Lambda,m_s,\mu_s,r)+\frac{\mu^2_s}{8m^2_q}H_1(\Lambda,m_s,\mu_s,r)+\frac{\mu^2_s}{2m^2_q}H_2(\Lambda,m_s,\mu_s,r)\mathbf{L}\cdot\mathbf{S}_{ij}\right]
\end{equation}
Here $\mathbf{L}=\mathbf{r}\times\mathbf{p}$ is the angular momentum
operator.
\item{Vector boson exchange}
\begin{eqnarray}
\nonumber
V_v(\mathbf{r})&=&\frac{\mu_v}{4\pi}\bigg\{g^2_{vqq}H_0(\Lambda,m_v,\mu_v,r)-\frac{(g^2_{vqq}+4g_{vqq}f_{vqq})\mu^2_v}{8m^2_q}H_1(\Lambda,m_v,\mu_v,r)\\
\nonumber&&-(g_{vqq}+f_{vqq})^2\frac{\mu^2_v}{12m^2_q}\Big[H_3(\Lambda,m_v,\mu_v,r)S_{ij}(\hat{\mathbf{r}})+2H_1(\Lambda,m_v,\mu_v,r)(\bm{\sigma_i\cdot\bm{\sigma}_j})\Big]\\
\label{11}&&-(3g^2_{vqq}+4g_{vqq}f_{vqq})\frac{\mu^2_{v}}{2m^2_q}H_2(\Lambda,m_v,\mu_v,r)\mathbf{L}\cdot\mathbf{S}_{ij}\bigg\}
\end{eqnarray}
For $I=1$ isovector boson exchange, the above potential should be
multiplied by the operator $\bm{\tau}_i\cdot\bm{\tau}_j$ in the
isospin space. We have included the contribution of the "$\delta$
function" term in the above potentials, which gives the delta
function when no regularization is used, since this contribution
turns out to be important \cite{Thomas:2008ja}. The effective
potential between two hadrons are obtained by summing the
interactions between light quarks or antiquarks via one boson
exchange.
\end{enumerate}

\section{Meson parameters and coupling constants}

As the well-known nuclear-nuclear interaction in the one boson
exchange model, we shall take into account the contributions from
pseudoscalar mesons $\pi$ and $\eta$ exchange, that from scalar
meson $\sigma$ exchange, and those from vector mesons $\rho$ and
$\omega$ exchange. The basic input parameters are the boson masses
and the effective coupling constants between the exchanged bosons
and the constituent quarks. The meson masses with their quantum
numbers are taken from the compilation of the Particle Data Group
\cite{pdg}. For the constituent quark-meson coupling constants, one
may derive suitable estimates from the phenomenologically known $\pi
NN$, $\eta NN$, $\sigma NN$, $\rho NN$ and $\omega NN$ coupling
constants using the Goldberger-Treiman relation. Riska and Brown
have demonstrated that the nucleon resonance transition couplings to
$\pi$, $\rho$ and $\omega$ can be derived in the single-quark
operator approximation \cite{Riska:2000gd}, which are in good
agreement with the experimental data. Along the same way, we can
straightforwardly derive the following relations between the
boson-quark couplings and the boson-nucleon couplings,
\begin{eqnarray}
\nonumber&&g_{\pi qq}=\frac{3}{5} \frac{m_q}{m_N}\,g_{\pi
NN},~~~~g_{\eta qq}=\frac{m_q}{m_{N}}\,g_{\eta NN}\\
\nonumber&&g_{\rho qq}=g_{\rho NN},~~~~~~~~~~~~f_{\rho
qq}=\frac{3}{5}\frac{m_q}{m_{N}}f_{\rho
NN}-(1-\frac{3}{5}\frac{m_q}{m_N})g_{\rho NN}\\
\nonumber&&g_{\omega qq}=\frac{1}{3}\,g_{\omega
NN},~~~~~~~~f_{\omega qq}=\frac{m_q}{m_N}f_{\omega
NN}-(\frac{1}{3}-\frac{m_q}{m_N})g_{\omega NN}\\
\label{12}&&g_{\sigma qq}=\frac{1}{3}\,g_{\sigma NN}
\end{eqnarray}
where $m_N$ is the nucleon mass. In the present work, the
constituent up(down) quark mass $m_{u(d)}$ is taken to be usual
value $m_{u(d)}\simeq313$ MeV, which is about one third of the
nucleon mass. The effective boson-nucleon coupling constants are
taken from the well-known Bonn model \cite{Machleidt:1987hj}, and a
typical set of parameters is shown in Table \ref{parameter}. The
uncertainty of the effective couplings will be taken into account
later, all the coupling constants except $g_{\pi NN}$ would be
reduced by a factor of two, since the experimental value for $g_{\pi
NN}$ has been determined accurately from pion-nucleon and
nucleon-nucleon scatterings. In the following, we shall explore the
possible molecular states consisting of a pair heavy flavor
pseudoscalar and vector mesons, their masses are taken from Particle
Data Group \cite{pdg}: ${m_{D^{0}}=1864.84}$ MeV, ${
m_{D^{\pm}}=1869.62}$ MeV, ${m_{D^{*0}}=2006.97}$ MeV,
${m_{D^{*\pm}}=2010.27}$ MeV, ${m_{B^{0}}=5279.53}$ MeV, ${
m_{B^{\pm}}=5279.15}$ MeV and ${m_{B^{*}}=5325.1}$ MeV.

\begin{table}[hptb]
\begin{center}
\begin{tabular}{|ccccc|}\hline\hline
Boson &~~~$I^{G}(J^P)$ &~~~Mass (MeV) & ~~~$g^2/4\pi$  &
~~~$f^2/4\pi$
\\\hline
$\pi^{\pm}$ & $1^-(0^-)$   & 139.57  &  14.9 &     \\
$\pi^0$  &  $1^-(0^-)$  &  134.98   &14.9 &     \\
$\eta$  & $0^+(0^-)$ &  547.85     &3.0 &    \\
$\sigma$  &  $0^+(0^+)$ & 600.0      &  7.78 &   \\
$\rho$ &  $1^+(1^-)$  & 775.49 &   0.95  & 35.35\\
$\omega$  &  $0^-(1^-)$   &782.65  & 20.0 & 0.0\\\hline\hline
\end{tabular}
\end{center}
\caption{\label{parameter}Spin, parity, isospin, G-parity, the
masses of the exchange bosons, and the meson-nucleon coupling
constants in the model.}
\end{table}

\section{Deuteron from one boson exchange model}
\begin{figure}[hptb]
\includegraphics[scale=.645]{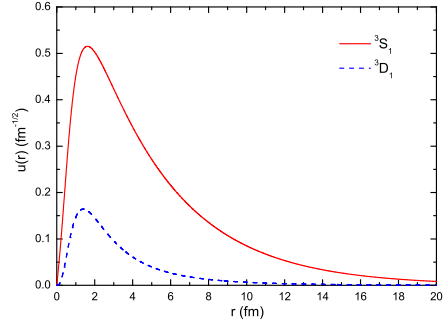}
\caption{The deuteron ${\rm ^3S_1}$ and ${\rm ^3D_1}$ wavefunction
with binding energy $\varepsilon_d\simeq 2.25$ MeV and
$\Lambda\simeq 808$ MeV. } \label{deuteron_wavefunction}
\end{figure}

\begin{figure}[hptb]
\begin{center}
\begin{tabular}{cc}
\includegraphics[scale=.645]{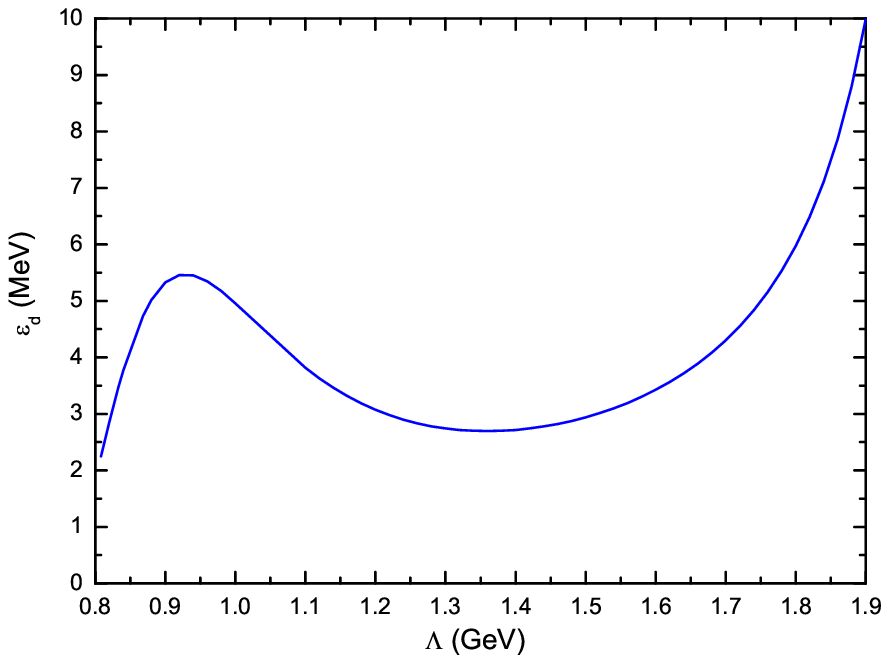}&\includegraphics[scale=.645]{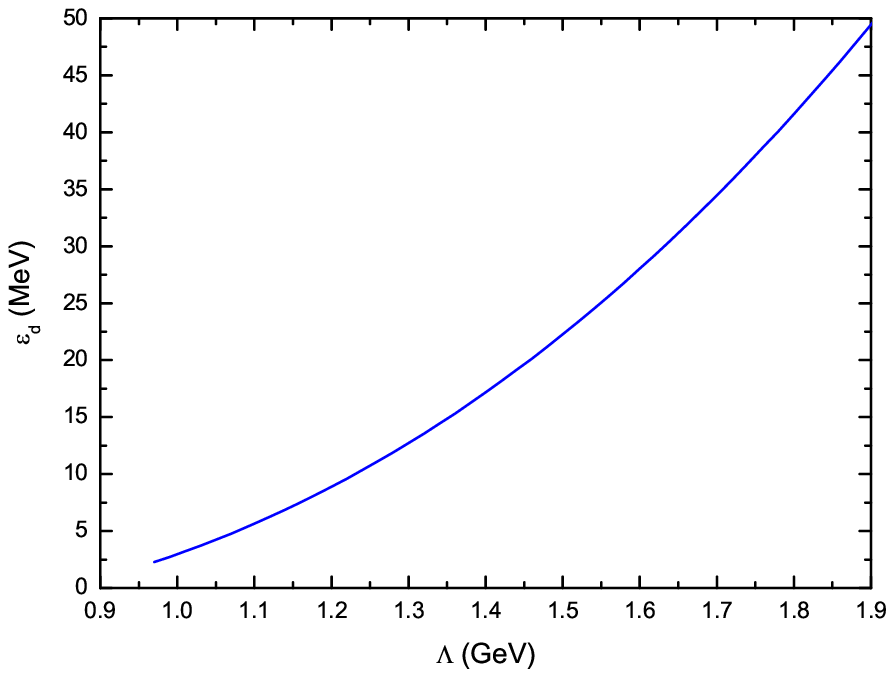}\\
(a)&(b)
\end{tabular}
\caption{\label{deuteron_binding_energy_variation}The deuteron
binding energy variation with respect to the regularization
parameter $\Lambda$. Fig. \ref{deuteron_binding_energy_variation}a
corresponds to the coupling constants shown in Table
\ref{parameter}, and Fig. \ref{deuteron_binding_energy_variation}b
for the couplings reduced by half.}
\end{center}
\end{figure}

\begin{figure}[hptb]
\begin{center}
\begin{tabular}{ccc}
\includegraphics[scale=.55]{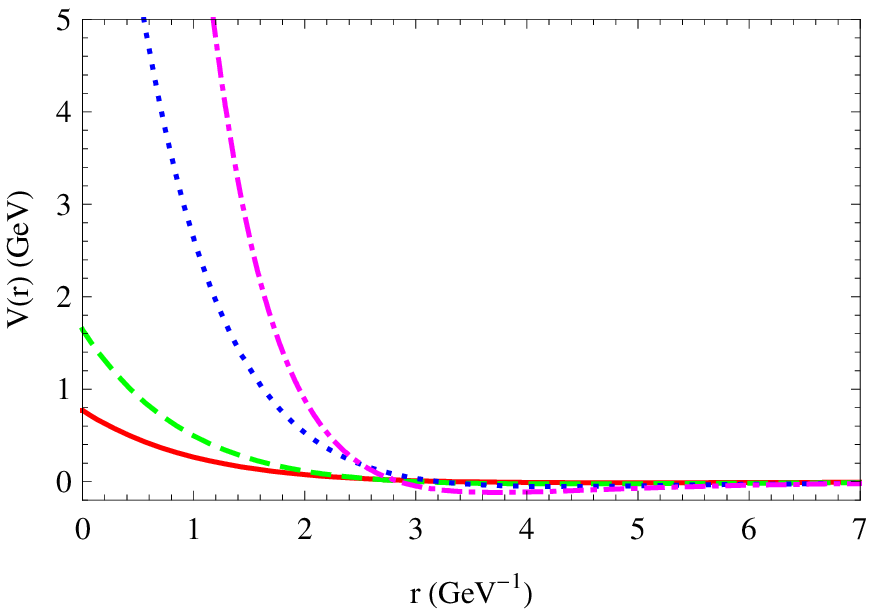}&\includegraphics[scale=.55]{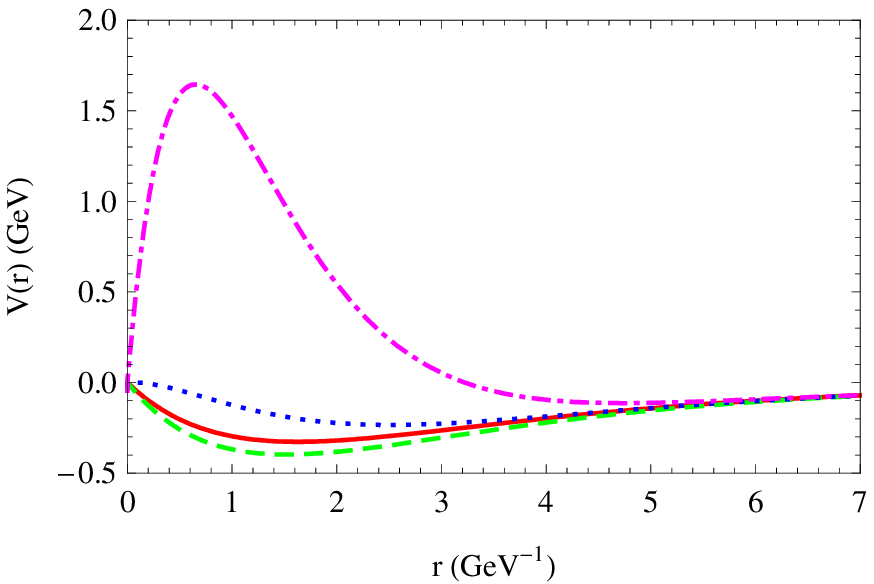}&\includegraphics[scale=.55]{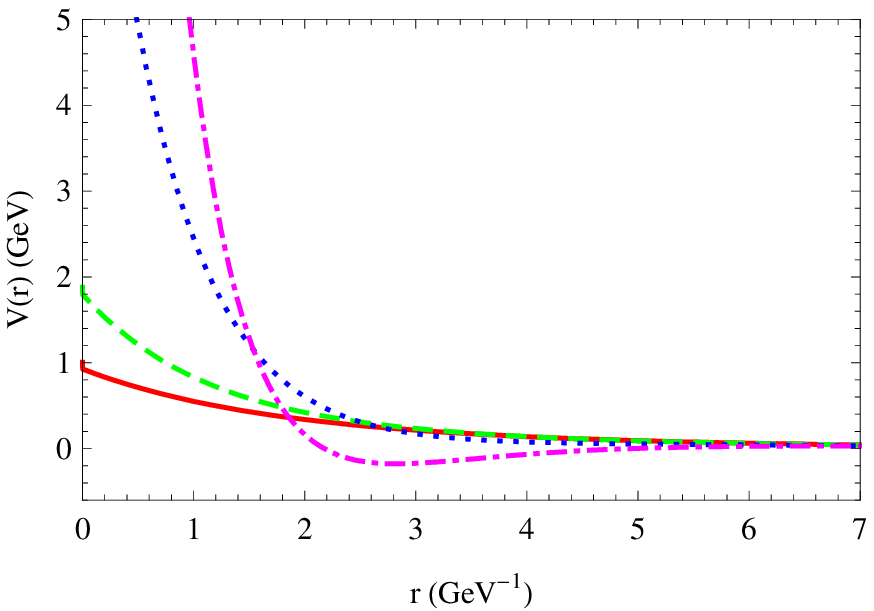}\\
(a)&(b)&(c)
\end{tabular}
\caption{\label{deuteron_effective_potential} The three components
of the deuteron effective potential in Eq.(\ref{13}), Fig.
\ref{deuteron_effective_potential} a, Fig.
\ref{deuteron_effective_potential} b and Fig.
\ref{deuteron_effective_potential} c respectively illustrate the
$V_{11}(\Lambda,r)$, $V_{12}(\Lambda,r)$ and $V_{22}(\Lambda,r)$
components. The solid line, dashed, dotted and dash-dotted lines
correspond to $\Lambda=0.8$ GeV, 0.9 GeV, 1.2 GeV and 1.6 GeV
respectively.}
\end{center}
\end{figure}

Deuteron is a uncontroversial proton-neutron bound state with $J=1$
and $I=0$. It has been established that the long distance one pion
exchange is the main binding mechanism, and the tensor force plays a
crucial role, which results in the ${\rm ^3S_1}$ and ${\rm ^3D_1}$
states mixing. Tornqvist and Close only considered the pion exchange
contribution in Refs. \cite{Tornqvist:1993ng,Thomas:2008ja},
however, the scalar meson $\sigma$ exchange and the vector mesons
$\rho$, $\omega$ exchange turn out to be important in providing the
short distance repulsion and the intermediate range attraction,
consequently, we shall take into account the contributions from the
heavier boson exchange in the following. The effective potential
becomes
\begin{eqnarray}
\nonumber V^{d}(\mathbf{r})&=&V^d_{\pi}(\mathbf{r})+V^d_{\eta}(\mathbf{r})+V^d_{\sigma}(\mathbf{r})+V^d_{\rho}(\mathbf{r})+V^d_{\omega}(\mathbf{r})\\
\nonumber&\equiv&
V^{d}_C(r)+V^d_S(r)(\bm{\sigma}_1\cdot\bm{\sigma}_2)+V^d_I(r)(\bm{\tau}_1\cdot\bm{\tau}_2)+V^d_T(r)S_{12}(\mathbf{\hat{r}})+
V^d_{SI}(r)(\bm{\sigma}_1\cdot\bm{\sigma}_2)(\bm{\tau}_1\cdot\bm{\tau}_2)\\
\label{13}&&+V^d_{TI}(r)S_{12}(\mathbf{\hat{r}})(\bm{\tau}_1\cdot\bm{\tau}_2)+V^d_{LS}(r)(\mathbf{L}\cdot\mathbf{S})+V^d_{LSI}(r)(\mathbf{L}\cdot\mathbf{S})(\bm{\tau}_1\cdot\bm{\tau}_2)
\end{eqnarray}
where $\mathbf{S}=\frac{1}{2}(\bm{\sigma}_1+\bm{\sigma}_2)$ is the
total spin, and $\mathbf{L}$ is the relative angular momentum
operator. In the isospin symmetry limit, the components
$V^{d}_C(r)$, $V^d_S(r)$ etc are given by
\begin{eqnarray}
\nonumber V^d_C(r)&=&-\frac{g^2_{\sigma
NN}}{4\pi}\,m_{\sigma}\Big[H_0(\Lambda,m_{\sigma},m_{\sigma},r)+\frac{m^2_{\sigma}}{8m^2_N}H_1(\Lambda,m_{\sigma},m_{\sigma},r)\Big]+\frac{g^2_{\omega
NN}}{4\pi}\,m_{\omega}H_0(\Lambda,m_{\omega},m_{\omega},r)\\
\nonumber&&-\frac{g^2_{\omega NN}+4g_{\omega NN}f_{\omega
NN}}{4\pi}\frac{m^3_{\omega}}{8m^2_N}H_1(\Lambda,m_{\omega},m_{\omega},r)\\
\nonumber V^d_S(r)&=&-\frac{g^2_{\eta
NN}}{4\pi}\frac{m^3_{\eta}}{12m^2_N}H_1(\Lambda,m_{\eta},m_{\eta},r)-\frac{(g_{\omega
NN}+f_{\omega
NN})^2}{4\pi}\frac{m^3_{\omega}}{6m^2_{N}}H_1(\Lambda,m_{\omega},m_{\omega},r)\\
\nonumber V^d_I(r)&=&\frac{g^2_{\rho
NN}}{4\pi}m_{\rho}H_0(\Lambda,m_{\rho},m_{\rho},r)-\frac{g^2_{\rho
NN}+4g_{\rho NN}f_{\rho
NN}}{4\pi}\frac{m^3_{\rho}}{8m^2_{N}}H_1(\Lambda,m_{\rho},m_{\rho},r)\\
\nonumber V^{d}_T(r)&=&\frac{g^2_{\eta
NN}}{4\pi}\frac{m^3_{\eta}}{12m^2_{N}}H_3(\Lambda,m_{\eta},m_{\eta},r)-\frac{(g_{\omega
NN}+f_{\omega
NN})^2}{4\pi}\frac{m^3_{\omega}}{12m^2_{N}}H_3(\Lambda,m_{\omega},m_{\omega},r)\\
\nonumber V^{d}_{SI}(r)&=&-\frac{g^2_{\pi
NN}}{4\pi}\frac{m^3_{\pi}}{12m^2_{N}}H_1(\Lambda,m_{\pi},m_{\pi},r)-\frac{(g_{\rho
NN}+f_{\rho
NN})^2}{4\pi}\frac{m^3_{\rho}}{6m^2_{N}}H_1(\Lambda,m_{\rho},m_{\rho},r)\\
\nonumber V^{d}_{TI}(r)&=&\frac{g^2_{\pi
NN}}{4\pi}\frac{m^3_{\pi}}{12m^2_{N}}H_3(\Lambda,m_{\pi},m_{\pi},r)-\frac{(g_{\rho
NN}+f_{\rho
NN})^2}{4\pi}\frac{m^3_{\rho}}{12m^2_{N}}H_3(\Lambda,m_{\rho},m_{\rho},r)\\
\nonumber V^{d}_{LS}(r)&=&-\frac{g^2_{\sigma
NN}}{4\pi}\frac{m^3_{\sigma}}{2m^2_{N}}H_2(\Lambda,m_{\sigma},m_{\sigma},r)-\frac{3g^2_{\omega
NN}+4g_{\omega NN}f_{\omega
NN}}{4\pi}\frac{m^3_{\omega}}{2m^2_{N}}H_2(\Lambda,m_{\omega},m_{\omega},r)\\
\label{14} V^{d}_{LSI}(r)&=&-\frac{3g^2_{\rho NN}+4g_{\rho
NN}f_{\rho
NN}}{4\pi}\frac{m^3_{\rho}}{2m^2_{N}}H_2(\Lambda,m_{\rho},m_{\rho},r)
\end{eqnarray}
In the basis of ${\rm ^3S_1}$ and ${\rm ^3D_1}$ states, the deuteron
potential can be written in the matrix form as
\begin{eqnarray}
\nonumber
V^d&=&\big[V^d_C(r)+V^d_S(r)-3V^d_I(r)-3V^d_{SI}(r)\big]\left(\begin{array}{cc}1&0\\
0&1\end{array}\right)+\big[9V^d_{LSI}(r)-3V^d_{LS}(r)\big]\left(\begin{array}{cc}0&0\\
0&1\end{array}\right)\\
\label{15}&&+\big[V^d_{T}(r)-3V^d_{TI}(r)\big]\left(\begin{array}{cc}0&\sqrt{8}\\
\sqrt{8}&-2\end{array}\right)
\end{eqnarray}
Taking into account the centrifugal barrier from D wave and solving
the corresponding two channel Schr$\ddot{\rm o}$dinger equation
numerically via the Fortran77 package FESSDE2.2 \cite{fessde}, which
can fastly and accurately solve the eigenvalue problem for systems
of coupled Schr$\ddot{\rm o}$dinger equations, we find the binding
energy $\varepsilon_d\simeq 2.25$ MeV for the cutoff parameter
$\Lambda=808$ MeV, and the corresponding wavefunction is presented
in Fig. \ref{deuteron_wavefunction}. If we reduce half of the
effective coupling constants except $g_{\pi NN}$, the binding energy
is found to be about 2.28 MeV with $\Lambda=970$ MeV. From the
wavefunction one can calculate the static properties of deuteron
such as the root of mean square radius, the D wave probability, the
magnetic moment and the quadrupole moment, which are in agreement
with experimental data. We would like to note that the small binding
energy of deuteron is a cancellation result of different
contributions of opposite signs. The detailed results are listed in
Table \ref{deuteron-static-properties}, it is obvious the results
are sensitive to the regularization parameter $\Lambda$, and the
same conclusion has been drawn in the one pion exchange model
\cite{Tornqvist:1993ng,Thomas:2008ja}. The binding energy variation
with respect to $\Lambda$ is shown in Fig.
\ref{deuteron_binding_energy_variation}, the dependence is less
sensitive than the one pion exchange model. It is obvious that the
binding energy variation with $\Lambda$ is dependent on the coupling
constants. For the coupling constants listed in Table
\ref{parameter}, the binding energy no longer monotonically
increases with $\Lambda$ in contrast with the one pion exchange
model. To understand this peculiar behavior, we plot the three
components of the deuteron effective potential in Eq.(\ref{15}) in
Fig. \ref{deuteron_effective_potential}. We can see that both
$V_{11}(\Lambda,r)$ and $V_{22}(\Lambda,r)$ potentials are
repulsive, and they increase with $\Lambda$ at short distance.
However, at intermediate distance the relation
$|V_{12}(\Lambda=1.2\rm {GeV},r)|<|V_{12}(\Lambda=0.8\rm
{GeV},r)|<|V_{12}(\Lambda=0.9\rm {GeV},r)|<|V_{12}(\Lambda=1.6\rm
{GeV},r)|$ is satisfied, the $V_{12}(\Lambda,r)$  doesn't
monotonically increases with $\Lambda$. Therefore the non-monotonous
behavior in Fig. \ref{deuteron_binding_energy_variation}a mainly
comes from the non-monotonous dependence of $V_{12}(\Lambda,r)$
potential on $\Lambda$, which is a cancellation result of various
contributions. As has been discussed above, the heavy flavor system
should admit a larger $\Lambda$ than the deuteron. Therefore the
above values of $\Lambda$ with which the smaller deuteron binding
energy is reproduced, would be assumed to be the lower bound in the
following.

\begin{center}
\begin{table}[hptb]
\begin{tabular}{|cccccc|}\hline\hline
$\Lambda({\rm MeV})$&$~~~{\rm \varepsilon_d}(\rm MeV)$&$~~~{\rm
r}_{\rm rms}({\rm fm})$&$~~~{\rm P_D:P_S}(\%)$& $~~~~~{\rm
\mu_d}(\mu_N)$ &~~~${\rm Q_d(fm^2)}$\\\hline

808&2.25 &3.85 & 5.66:94.34& 0.85&0.27 \\

900&5.33&2.77&  7.44:92.56 &0.84&0.20\\

1000&4.96&2.87 &7.37:92.63& 0.84&0.21\\
\hline\hline

\multicolumn{6}{|c|}{all couplings are reduced by half except
$g_{\pi NN}$ }\\\hline\hline

$\Lambda({\rm MeV})$&$~~~{\rm \varepsilon_d}(\rm MeV)$&$~~~{\rm
r}_{\rm rms}({\rm fm})$&$~~~{\rm P_D:P_S}$& $~~~~~{\rm
\mu_d}(\mu_N)$ &~~~${\rm Q_d(fm^2)}$\\\hline

970 &2.28 &3.84 &  6.52:93.48&0.84&0.28 \\

1100& 5.65 & 2.70& 8.92:91.08&0.83&0.20 \\

1200&8.89 &2.28   & 10.26:89.74   & 0.82  &0.16\\
\hline\hline

\end{tabular}
\caption{\label{deuteron-static-properties}The deuteron static
properties in the one boson exchange potential model, where
$\varepsilon_d$ is the binding energy, ${\rm r_{rms}}$ is the root
of mean square radius, ${\rm P_S}$ and ${\rm P_D }$ represent the
S-state and D-state probabilities respectively, ${\rm \mu_d}$ is the
magnetic moment, and ${\rm Q_d}$ denotes the quadrupole moment.}
\end{table}
\end{center}

\section{Possible ${D\bar{D}^{*}/\bar{D}D^{*}}$  hadronic molecule and X(3872) }

The narrow charmoniumlike state X(3872) was discovered by the Belle
collaboration in the decay $B^{+}\rightarrow K^{+}+X(3872)$ followed
by $X(3872)\rightarrow J/\psi\pi^{+}\pi^{-}$ with a statistical
significance of 10.3$\sigma$ \cite{Choi:2003ue}. The existence of
X(3872) has been confirmed by CDF \cite{Acosta:2003zx}, D0
\cite{Abazov:2004kp} and Babar collaboration \cite{Aubert:2004ns}.
the CDF collaboration measured the X(3872) mass to be ${\rm
(3871.61\pm0.16(stat)\pm0.19(sys.))}$ MeV. Its quantum number is
strongly preferred to be $1^{++}$ \cite{Abulencia:2006ma}. In the
one pion exchange model, Tornqvist suggested that X(3872) is a
$1^{++}$ $D\bar{D}^{*}/\bar{D}D^{*}$ molecule and isospin is
strongly broken \cite{Tornqvist:2004qy}. Recently Close et al.
re-analyzed X(3872) in the same model, the critical overall sign is
corrected and the contribution of the "$\delta$ function" term is
included \cite{Thomas:2008ja}. Swanson have taken into account both
the long rang pion exchange and short range contribution arising
from constituent quark interchange \cite{swanson}. Recently Zhu et
al. dynamically studied the binding of X(3872) in the heavy quark
effective theory \cite{Liu:2008fh}. In this section, we will
investigate the $1^{++}$ $D\bar{D}^{*}/\bar{D}D^{*}$ system from the
one boson exchange model at quark level, where the short range
interactions are represented by the heavier bosons $\eta$, $\sigma$,
$\rho$ and $\omega$ exchange instead of the quark interchange.

\begin{figure}
\includegraphics[scale=.645]{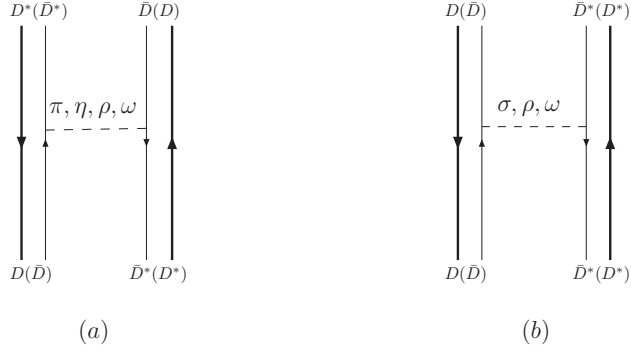}
\caption{$D\bar{D}^{*}$ and $\bar{D}D^{*}$ interaction in one boson
exchange model at quark level, where the thick line represents heavy
quark or antiquark, and the thin line denotes light quark or
antiquark.} \label{interaction diagram}
\end{figure}

There is only a sign difference $(-1)^{G}$ between the quark-quark
interaction and quark-antiquark interaction, and the magnitudes are
the same, where $G$ is the $G$-parity of the exchanged meson. The
diagrams contributing to the $D\bar{D}^{*}$ and $\bar{D}D^{*}$
interactions are displayed in Fig. \ref{interaction diagram}.
Because of the parity conservation, $D\bar{D}^{*}$ can only scatter
into $D^{*}\bar{D}$ via the pseudoscalar $\pi$ and $\eta$ exchange,
and $D\bar{D}^{*}$ scatters into $D\bar{D}^{*}$ with the scalar
$\sigma$ exchange, whereas the vector mesons $\rho$ and $\omega$
exchange contribute to both processes. The effective potential for
the $1^{++}$ $D\bar{D}^{*}/\bar{D}D^{*}$ system is
\begin{eqnarray}
\nonumber V^{X}(\mathbf{r})&=&-V^{X}_{\pi}(\mathbf{r})+V^{X}_{\eta}(\mathbf{r})+V^{X}_{\sigma}(\mathbf{r})+V^{X}_{\rho}(\mathbf{r})-V^{X}_{\omega}(\mathbf{r})\\
\nonumber&\equiv&
V^{X}_C(r)+V^X_S(r)(\bm{\sigma}_i\cdot\bm{\sigma}_j)+V^X_I(r)(\bm{\tau}_i\cdot\bm{\tau}_j)+V^X_T(r)S_{ij}(\mathbf{\hat{r}})+
V^X_{SI}(\mu_k,r)(\bm{\sigma}_i\cdot\bm{\sigma}_j)(\bm{\tau}_i\cdot\bm{\tau}_j)\\
\label{16}&&+V^X_{TI}(\mu_k,r)S_{ij}(\mathbf{\hat{r}})(\bm{\tau}_i\cdot\bm{\tau}_j)+V^X_{LS}(r)(\mathbf{L}\cdot\mathbf{S}_{ij})+V^X_{LSI}(r)(\mathbf{L}\cdot\mathbf{S}_{ij})(\bm{\tau}_i\cdot\bm{\tau}_j)
\end{eqnarray}
with $i$ and $j$ is the index of light quark or antiquark,
$\mu_k(k=1,2,3,4)$ takes four different values due to the mass
difference within the $D$, $D^{*}$ and $\pi$ isospin multiplets. The
eight components $V^{X}_C(r)$, $V^X_S(r)$ etc are given by
\begin{eqnarray}
\nonumber V^{X}_C({r})&=&-\frac{g^2_{\sigma
qq}}{4\pi}m_{\sigma}\Big[H_0(\Lambda,m_{\sigma},m_{\sigma},r)+\frac{m^2_{\sigma}}{8m^2_q}H_1(\Lambda,m_{\sigma},m_{\sigma},r)\Big]-\frac{g^2_{\omega
qq}}{4\pi}m_{\omega}H_0(\Lambda,m_{\omega},m_{\omega},r)\\
\nonumber&&+\frac{g^2_{\omega qq}+4g_{\omega qq}f_{\omega
qq}}{4\pi}\frac{m^3_{\omega}}{8m^2_q}H_1(\Lambda,m_{\omega},m_{\omega},r)\\
\nonumber V^{X}_S({r})&=&-\frac{g^2_{\eta
qq}}{4\pi}\frac{\mu^3_5}{12m^2_q}H_1(\Lambda,m_{\eta},\mu_5,r)+\frac{(g_{\omega
qq}+f_{\omega
qq})^2}{4\pi}\frac{\mu^3_7}{6m^2_q}H_1(\Lambda,m_{\omega},\mu_7,r)\\
\nonumber V^{X}_I({r})&=&\frac{g^2_{\rho
qq}}{4\pi}m_{\rho}H_0(\Lambda,m_{\rho},m_{\rho},r)-\frac{g^2_{\rho
qq}+4g_{\rho qq}f_{\rho qq}}{4\pi}\frac{m^3_{\rho}}{8m^2_q}H_1(\Lambda,m_{\rho},m_{\rho},r)\\
\nonumber V^{X}_{T}({r})&=&\frac{g^2_{\eta
qq}}{4\pi}\frac{\mu^3_5}{12m^2_q}H_3(\Lambda,m_{\eta},\mu_5,r)+\frac{(g_{\omega
qq}+f_{\omega
qq})^2}{4\pi}\frac{\mu^3_7}{12m^2_q}H_3(\Lambda,m_{\omega},\mu_7,r)\\
\nonumber
V^{X}_{SI}(\mu,{r})&=&\left\{\begin{array}{cc}\frac{g^2_{\pi
qq}}{4\pi}\frac{\mu^3}{12m^2_{q}}H_1(\Lambda,m_{\pi^{\pm,0}},\mu,r)-\frac{(g_{\rho
qq}+f_{\rho
qq})^2}{4\pi}\frac{\mu^3_6}{6m^2_q}H_1(\Lambda,m_{\rho},\mu_6,r),&\mu^2>0\\
\frac{g^2_{\pi
qq}}{4\pi}\frac{\tilde{\mu}^3}{12m^2_{q}}G_1(\Lambda,m_{\pi^{\pm,0}},\tilde{\mu},r)-\frac{(g_{\rho
qq}+f_{\rho
qq})^2}{4\pi}\frac{\mu^3_6}{6m^2_q}H_1(\Lambda,m_{\rho},\mu_6,r),&~~~~~~\mu^2=-\tilde{\mu}^2<0
\end{array}\right.\\
\nonumber V^{X}_{TI}(\mu,r)&=&\left\{\begin{array}{cc}
-\frac{g^2_{\pi
qq}}{4\pi}\frac{\mu^3}{12m^2_q}H_3(\Lambda,m_{\pi^{\pm,0}},\mu,r)-\frac{(g_{\rho
qq}+f_{\rho
qq})^2}{4\pi}\frac{\mu^3_6}{12m^2_q}H_3(\Lambda,m_{\rho},\mu_6,r),&\mu^2>0\\
-\frac{g^2_{\pi
qq}}{4\pi}\frac{\tilde{\mu}^3}{12m^2_q}G_3(\Lambda,m_{\pi^{\pm,0}},\tilde{\mu},r)-\frac{(g_{\rho
qq}+f_{\rho
qq})^2}{4\pi}\frac{\mu^3_6}{12m^2_q}H_3(\Lambda,m_{\rho},\mu_6,r),&~~~~~~\mu^2=-\tilde{\mu}^2<0
\end{array}
\right.\\
\nonumber V^{X}_{LS}(r)&=&-\frac{g^2_{\sigma
qq}}{4\pi}\frac{m^3_{\sigma}}{2m^2_q}H_2(\Lambda,m_{\sigma},m_{\sigma},r)+\frac{3g^2_{\omega
qq}+4g_{\omega qq}f_{\omega
qq}}{4\pi}\frac{m^3_{\omega}}{2m^2_q}H_2(\Lambda,m_{\omega},m_{\omega},r)\\
\label{17} V^{X}_{LSI}(r)&=&-\frac{3g^2_{\rho qq}+4g_{\rho
qq}f_{\rho
qq}}{4\pi}\frac{m^3_{\rho}}{2m^2_q}H_2(\Lambda,m_{\rho},m_{\rho},r)
\end{eqnarray}
where
\begin{eqnarray}
\nonumber
\mu^2_1&=&m^2_{\pi^0}-(m_{D^{*0}}-m_{D^{0}})^2,~~~~~\mu^2_2=m^2_{\pi^{\pm}}-(m_{D^{*0}}-m_{D^{\pm}})^2\\
\nonumber
\mu^2_3&=&m^2_{\pi^{\pm}}-(m_{D^{*\pm}}-m_{D^{0}})^2,~~~~~\mu^2_4=m^2_{\pi^0}-(m_{D^{*\pm}}-m_{D^{\pm}})^2\\
\nonumber\mu^2_5&=&m^2_{\eta}-(m_{D^{*0}}-m_{D^{0}})^2,~~~~~~~\mu^2_6=m^2_{\rho}-(m_{D^{*0}}-m_{D^{0}})^2\\
\label{18}\mu^2_5&=&m^2_{\omega}-(m_{D^{*0}}-m_{D^{0}})^2
\end{eqnarray}

\begin{figure}[hptb]
\includegraphics[scale=.745]{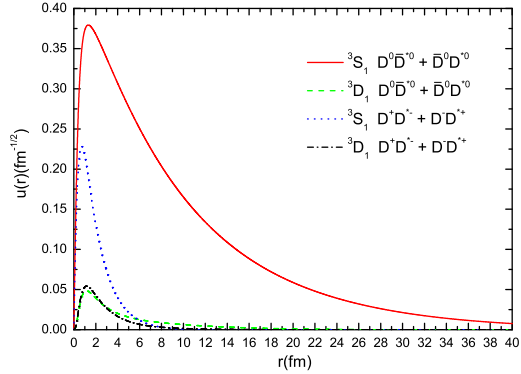}
\caption{\label{x3872 wavefunction} The four components spatial
wavefunctions of the $1^{++}$ $D\bar{D}^{*}/\bar{D}D^{*}$ system
with $\Lambda=808$ MeV. }
\end{figure}

These $\mu^2$ parameters approximately represent the recoil effect
due to different values of $m_{D}$ and $m_{D^{*}}$ as in Refs.
\cite{Tornqvist:1993ng,Thomas:2008ja}. For the $\eta$, $\sigma$,
$\rho$ and $\omega$ exchange processes, the mass difference of
$m_{D^{0}}$ and $m_{D^{\pm}}$ as well as $m_{D^{*0}}$ and
$m_{D^{*\pm}}$ are neglected, since they are much smaller comparing
with $m_{\eta}$, $m_{\rho}$ and $m_{\omega}$. X(3872) is very close
to the $D^{0}\bar{D}^{*0}$ threshold, however, it is about 8.3 MeV
below the $D^{+}D^{*-}$ threshold. Hence, isospin symmetry is
drastically broken \cite{Close:2003sg,Tornqvist:2004qy}. For the
$J^{PC}=1^{++}$ $D\bar{D}^{*}/\bar{D}D^{*}$ system, they can be in S
wave or D wave similar to the deuteron, then the wavefunction of
this system is written as
\begin{eqnarray}
\nonumber |X(3872)\rangle&=&\frac{u_1(r)}{r}\frac{1}{\sqrt{2}}|(D^{0}\bar{D}^{*0}+\bar{D}^{0}D^{*0})_S\rangle+\frac{u_2(r)}{r}\frac{1}{\sqrt{2}}|(D^{0}\bar{D}^{*0}+\bar{D}^{0}D^{*0})_D\rangle\\
\label{19}&&+\frac{u_3(r)}{r}\frac{1}{\sqrt{2}}|(D^{+}D^{*-}+D^{-}D^{*+})_S\rangle+\frac{u_4(r)}{r}\frac{1}{\sqrt{2}}|(D^{+}D^{*-}+D^{-}D^{*+})_D\rangle
\end{eqnarray}
where the subscript $S$ and $D$ denote the system in $S$ wave and
$D$ wave respectively. $u_1(r)$, $u_2(r)$, $u_3(r)$ and $u_4(r)$ are
the spatial wavefunctions. There are four channels coupled with each
other as has been shown above, and we might as well choose the basis
to be
$|1\rangle\equiv\frac{1}{\sqrt{2}}|(D^{0}\bar{D}^{*0}+\bar{D}^{0}D^{*0})_S\rangle$,
$|2\rangle\equiv\frac{1}{\sqrt{2}}|(D^{0}\bar{D}^{*0}+\bar{D}^{0}D^{*0})_D\rangle$,
$|3\rangle\equiv\frac{1}{\sqrt{2}}|(D^{+}D^{*-}+D^{-}D^{*+})_S\rangle$
and
$|4\rangle\equiv\frac{1}{\sqrt{2}}|(D^{+}D^{*-}+D^{-}D^{*+})_D\rangle$.
Using the analytical formula for the matrix elements presented in
the appendix, the effective potential for $1^{++}$
$D\bar{D}^{*}/\bar{D}D^{*}$ can be written in the matrix form as
\begin{eqnarray}
\nonumber
V^{X}(r)&=&\big[V^{X}_C(r)+V^{X}_S(r)\big]\left(\begin{array}{cccc}
1&0&0&0\\
0&1&0&0\\
0&0&1&0\\
0&0&0&1
\end{array}\right)+\big[V^{X}_I(r)+V^{X}_{SI}(\mu_k,r)\big]\left(\begin{array}{cccc}
-1&0&-2&0\\
0&-1&0&-2\\
-2&0&-1&0\\
0&-2&0&-1
\end{array}\right)\\
\nonumber&&+V^{X}_{T}(r)\left(\begin{array}{cccc}
0&-\sqrt{2}&0&0\\
-\sqrt{2}&1&0&0\\
0&0&0&-\sqrt{2}\\
0&0&-\sqrt{2}&1
\end{array}\right)+V^{X}_{TI}(\mu_k,r)\left(\begin{array}{cccc}
0&\sqrt{2}&0&2\sqrt{2}\\
\sqrt{2}&-1&2\sqrt{2}&-2\\
0&2\sqrt{2}&0&\sqrt{2}\\
2\sqrt{2}&-2&\sqrt{2}&-1
\end{array}\right)\\
\label{20}&&+V^{X}_{LS}(r)\left(\begin{array}{cccc}
0&0&0&0\\
0&-3/2&0&0\\
0&0&0&0\\
0&0&0&-3/2
\end{array}\right)+V^{X}_{LSI}(r)\left(\begin{array}{cccc}
0&0&0&0\\
0&3/2&0&3\\
0&0&0&0\\
0&3&0&3/2
\end{array}\right)
\end{eqnarray}
In the above equation, the value of $\mu^2_k$ is $\mu^2_1$ for the
up-left $2\times2$ matrix elements, and it is equal to $\mu^2_4$ for
the down-right $2\times2$ matrix elements. There is ambiguity in
choosing $\mu^2_k$ value for the processes
$D^{0}\bar{D}^{*0}\rightarrow D^{*+}D^-$ or $D^{+}D^{*-}\rightarrow
D^{*0}\bar{D}^{0}$, accordingly $\mu^2_k$ can take the value
$\mu^2_2$ or $\mu^2_3$ for the off-diagonal $2\times2$ matrix
elements, the numerical results for both choices would be given in
the following. The different $\mu_k$ values is due to the isospin
symmetry breaking from the mass difference within the $D$, $D^{*}$
and $\pi$ isospin multiplets. Taking into account the centrifugal
barrier from D wave and solving the four channel coupled
Schr$\ddot{\rm o}$dinger equation using the package FESSDE2.2, the
numerical results are listed in Table \ref{x3872}. It is remarkable
that the $1^{++}$ $D\bar{D}^{*}/\bar{D}D^{*}$ system could
accomodate a molecular state with mass about 3871.6 MeV for
$\Lambda=808$ MeV, it is very close to the central value of X(3872)
mass 3871.61 MeV. The corresponding wavefunction is shown in Fig.
\ref{x3872 wavefunction}, it is obvious that the
$D^{0}\bar{D}^{*0}+\bar{D}^{0}D^{*0}$ component dominates over the
$D^{+}D^{*-}+D^{-}D^{*+}$ component. Since the spatial wavefunctions
$u_1(r)$ and $u_3(r)$ have the same sign, the same is true for
$u_2(r)$ and $u_4(r)$, thus the $I=0$ component in this state is
predominant, it would be a isospin singlet in the isospin symmetry
limit. From the results in Table \ref{x3872}, we notice that the
predictions about the static properties for the two $\mu^2$ choices
are very similar to each other, and the difference is small. The
isospin symmetry is strongly broken especially for the states near
the threshold. The uncertainties induced by the effective coupling
constants are considered, we reduce half of the couplings except
$g_{\pi NN}$, and the numerical results are given in Table
\ref{x3872} as well. For both choices of the coupling constants, the
binding energy and other static properties are sensitive to the
regularization parameter $\Lambda$, and the bound state mass
dependence on $\Lambda$ is displayed in Fig. \ref{x3872 binding
energy variation}. It is obvious that the bound state mass decreases
monotonically with the regularization parameter $\Lambda$ as in the
one pion exchange model. In short summary, the predictions are
qualitatively the same as those in the one pion exchange model, even
after we have included the contributions from $\eta$, $\sigma$,
$\rho$ and $\omega$ exchange. Since unexpectedly large branch ratio
of $X(3872)\rightarrow\psi(2S)\gamma$ recently was reported
\cite{Fulsom:2008rn}, we have to take into account the mixing
between the $1^{++}$ $D\bar{D}^{*}/\bar{D}D^{*}$ molecule and the
conventional charmonium state in order to identify this state with
X(3872). This is outside the range of the present work.

\begin{center}
\begin{table}[hptb]
\begin{tabular}{|c|cccc|}\hline\hline


$\mu^2$ &$\Lambda({\rm MeV})$&$~~~{\rm M}(\rm MeV)$&$~~~{\rm r}_{\rm
rms}({\rm fm})$&$~~~{\rm
P^{00}_S:P^{00}_D:P^{+-}_S:P^{+-}_D}(\%)$\\\hline

 & 808& 3871.6  & 7.02 &  90.76:0.56:8.11:0.56\\

 & 840&  3870.4 &2.84 & 78.23:1.08:19.59:1.11 \\

$\mu^2_3$ &850&3869.8 & 2.45& 75.26:1.21:22.29:1.23 \\

 &900& 3865.9 & 1.61 & 65.17:1.89:31.06:1.88 \\

 & 1000&  3849.2 & 1.08 &  53.06:4.72:37.65:4.57\\\hline

 & 808&  3871.7& 11.34 & 94.40:0.38:4.86:0.36 \\

 & 840 & 3870.7&  3.19&  80.74:0.99:17.26:1.01\\

$\mu^2_2$ & 850 & 3870.2 & 2.68 & 77.44:1.14:20.28:1.15 \\

 & 900& 3866.4&1.66 & 66.23:1.85:30.09:1.83\\

& 1000& 3849.9 &  1.09& 53.35:4.69:37.43:4.53 \\
\hline\hline

\multicolumn{5}{|c|}{all couplings except $g_{\pi NN}$ are reduced
by half }\\\hline\hline

$\mu^2$& $\Lambda({\rm MeV})$&$~~~{\rm M}(\rm MeV)$&$~~~{\rm r}_{\rm
rms}({\rm fm})$&$~~~{\rm P^{00}_S:P^{00}_D:P^{+-}_S:P^{+-}_D}(\%)$\\
\hline

 &970& 3869.1& 2.13& 70.65:1.65:26.02:1.69
\\

$\mu^2_3$ &1100& 3860.1& 1.25&57.24:2.98:36.83:2.95
\\

 &1200& 3848.2 &1.00 & 51.80:4.40:39.46:4.33
\\
\hline

  &970&3869.5 &2.28 & 72.56:1.57:24.28:1.60
\\

$\mu^2_2$ &1100&3860.8& 1.27 &57.76:2.94:36.38:2.92
\\

 &1200& 3849.0&  1.01 & 52.04:4.38:39.29:4.30
\\
\hline\hline

\end{tabular}
\caption{\label{x3872}The predictions about the mass, the root of
mean square radius(rms) and the probabilities of the different
components for the $1^{++}$ $D\bar{D}^{*}/\bar{D}D^{*}$ molecule.}
\end{table}
\end{center}


\begin{figure}[hptb]
\begin{center}
\begin{tabular}{cc}
\includegraphics[scale=.645]{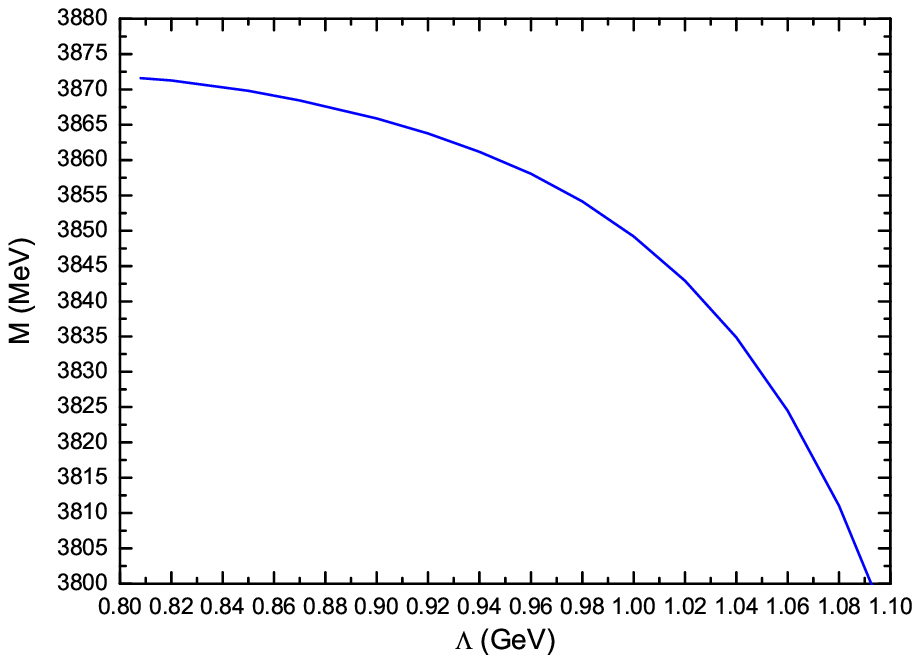}&\includegraphics[scale=.645]{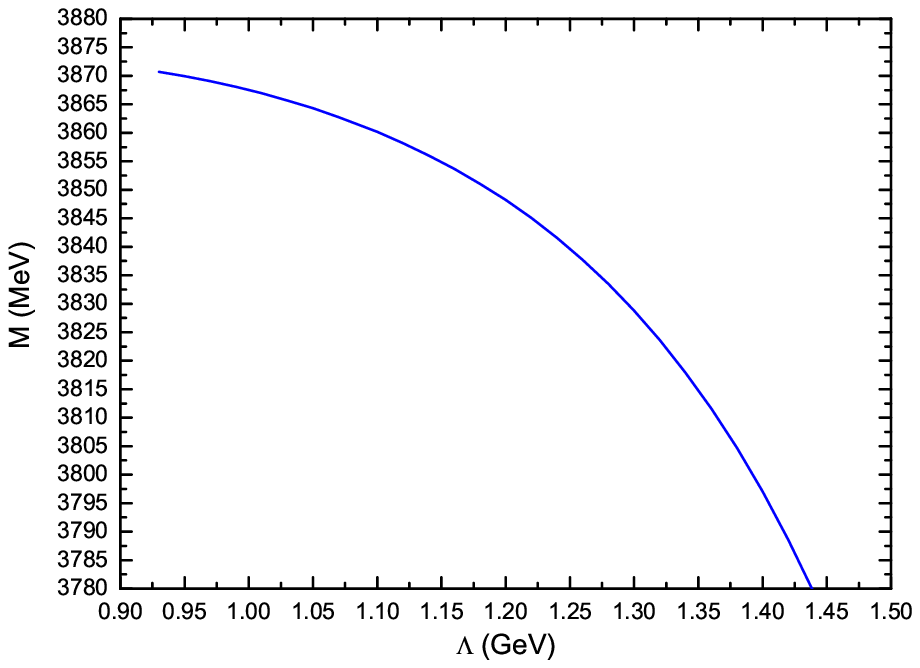}\\
(a)&(b)
\end{tabular}
\caption{\label{x3872 binding energy variation}The variation of the
$1^{++}$ $D\bar{D}^{*}/\bar{D}D^{*}$ bound state mass with respect
to $\Lambda$. (a) corresponds to the coupling constants shown in
Table \ref{parameter}, and (b) for the couplings reduced by half.}
\end{center}
\end{figure}

\section{Possible molecular states of other heavy flavor PV systems}
\subsection{$B\bar{B}^{*}/\bar{B}B^{*}$ system}
For the $1^{++}$ $B\bar{B}^{*}/\bar{B}B^{*}$ system, the kinetic
energy is greatly reduced due to the heavier mass of $B$ meson, and
the interaction potential has features similar to those of the
$D\bar{D}^{*}/\bar{D}D^{*}$ system except that the former is deeper
than the latter. Therefore molecular states should be more easily
formed. Following the same procedure as the
$D\bar{D}^{*}/\bar{D}D^{*}$ case, the numerical results are shown in
Table \ref{bottom analog of x3872}, where the $\mu^2$ ambiguity is
considered. For the same value of $\Lambda$, the
$B\bar{B}^{*}/\bar{B}B^{*}$ system is really more strongly bound
than the $D\bar{D}^{*}/\bar{D}D^{*}$ system, its binding energy is a
few tens of MeV, and the same was predicted in the one pion exchange
model \cite{Tornqvist:1993ng,Thomas:2008ja} and in the models
\cite{Liu:2008fh,liu-chiral}. It is obvious that the predictions
about the static properties for the two $\mu^2$ choices are
approximately the same. We notice that the isospin symmetry breaking
is less stronger than the charm system, this is because the mass
difference of $B^{0}$ and $B^{+}$ is smaller than that of $D^{0}$
and $D^{+}$ as well as $D^{*0}$ and $D^{*+}$. It is notable that
there may be two molecular states for appropriate values of
$\Lambda$. The corresponding wavefunctions for $\Lambda=1000$ MeV
and $\mu^2=\mu^2_{bb1}\equiv
m^2_{\pi^{\pm}}-(m_{B^{*}}-m_{B^{0}})^2$ are displayed in Fig.
\ref{wavefunction of bottom analog}, the first state is tightly
bound, whereas the second is loosely bound.  We notice that the
first state is almost an isospin singlet, and the $I=0$ component is
dominant for the second state. This state can no longer be produced
through $B$ meson decay because of its large mass, and we have to
resort to hadron collider. We can search for this state at Tevatron
via $p\bar{p}\rightarrow\pi^{+}\pi^{-}\Upsilon(1S)$, and LHC is more
promising.
\subsection{$DD^{*}$ system with $C=2$}
The interaction potentials arise from the one boson exchange between
two antiquarks instead of a quark and antiquark pair, hence both the
$\pi$ exchange and $\omega$ exchange potentials have overall
opposite sign relative to the $D\bar{D}^{*}/\bar{D}D^{*}$ case. In
this case we have four coupled channels $(D^{+}D^{*0})_S$,
$(D^{+}D^{*0})_D$,  $(D^{0}D^{*+})_S$ and $(D^{0}D^{*+})_D$. The
numerical results are given in Table \ref{static properties of C=2
system}. For $\Lambda=808$ MeV or $\Lambda=970$ MeV, we find no
bound state. A bound state with mass about 3873.9 MeV appears for
$\Lambda=1600$ MeV (about 3873.1 MeV for $\Lambda=1900$ MeV if the
couplings except $g_{\pi NN}$ are reduced half), and the
corresponding wavefunction is shown in Fig. \ref{wavefunction of
C=2}. We notice that the wavefunctions of $^3S_1$ $D^{+}D^{*0}$ and
$D^{0}D^{*+}$ have opposite signs, the same is true for the $^3D_1$
$D^{+}D^{*0}$ and $D^{0}D^{*+}$ wavefunctions, therefore this state
would be a isospin singlet in the isospin symmetry limit. We notice
that the $^3D_1$ probability are much larger than $^3S_1$
probability for the state with $\Lambda=1800$ MeV, although there is
centrifugal barrier for the D wave state. Thus the S-D mixing effect
induced by the tensor force is especially crucial for this state. In
short, the bound state of the $DD^{*}$ system appears only for the
regularization parameter $\Lambda$ as large as 1600 MeV or 1900 MeV,
which is beyond the range of 0.8 to 1.5 GeV favored by the
nucleon-nucleon interaction. Moreover, the parameters that allow
X(3872) to emerge as a $D\bar{D}^{*}/\bar{D}D^{*}$ molecule exclude
the $DD^{*}$ bound state, as can be seen from the results in section
V. Consequently we tend to conclude that the $DD^{*}$ molecular
state may not exist.
\subsection{$BB^{*}$ system with $B=2$}
The situation is very similar to the $DD^{*}$ system except the
different mass of $D$ mesons and $B$ mesons, we list the numerical
results in Table \ref{static properties of B=2 system}. We find a
marginally bound state with mass 10603.9 MeV for $\Lambda=808$ MeV,
which is very close to the $BB^{*}$ threshold. Its binding energy is
much smaller than that of the $1^{++}$ $B\bar{B}^{*}/\bar{B}B^{*}$
system, however, the binding energy is less sensitive to $\Lambda$
than the latter case. Fig. \ref{wavefunction of B=2} displays the
wavefunction of the bound state solution with mass 10602.3 MeV and
$\Lambda=900$ MeV. It is obvious that the $B^{+}B^{*0}$ and
$B^{0}B^{*+}$ wavefunctions have the opposite sign, then the $I=0$
component is dominant in this state. If the couplings except $g_{\pi
NN}$ are reduced by half, a weakly bound state with mass about
10601.5 MeV is found as well assuming $\Lambda=970$ MeV. These
indicates a weakly bound $BB^{*}$ should exist, This is consistent
with the results of Manohar and Wise form heavy quark effective
theory \cite{manohar}. For a loosely bound molecule, the leading
source of decay is dissociation, to a good approximation the
dissociation will proceed via the free space decay of the
constituent mesons.  The spin-parity forbids its decay into $BB$,
therefore the $BB^{*}$ molecule is a very narrow state and it mainly
decays into $BB\gamma$.
\subsection{Pseudoscalar-vector system with $C=B=1$}
This system could have the same quantum as $B_c$ meson or its
antiparticle, and it is different from all the systems discussed
above, eight channels instead of four channels are coupled with each
other under the one boson exchange interaction, i.e.
$(D^{+}B^{*0})_S$, $(D^{+}B^{*0})_D$, $(D^{0}B^{*+})_S$,
$(D^{0}B^{*+})_D$, $(D^{*+}B^{0})_S$, $(D^{*+}B^{0})_D$,
$(D^{*0}B^{+})_S$ and $(D^{*0}B^{+})_D$. We can investigate the
possible bound states along the same line, although it is somewhat
lengthy and tedious. There is ambiguity in choosing the $\mu^2$
value as well, for the $DB^{*}\rightarrow D^{*}B$ scattering
process, we could take $\mu^2=m^2_{ex}-(m_{D^{*}}-m_{D})^2$ or
$\mu^2=m^2_{ex}-(m_{B^{*}}-m_{B})^2$, where $m_{ex}$ is the mass of
the exchanged boson. Specifically for $D^{+}B^{*0}\rightarrow
D^{*+}B^{0}$ via $\pi$ exchange, we can choose
$\mu^2=m^2_{\pi^0}-(m_{D^{*+}}-m_{D^{+}})^2$ or
$\mu^2=m^2_{\pi^0}-(m_{B^{*0}}-m_{B^{0}})^2$. This ambiguity has
been taken into account in our analysis. The numerical results are
given in Table \ref{static properties of C=B=1 system}, For
$\Lambda=808$ MeV, we find no bound state. With the choice
$\mu^2=m^2_{ex}-(m_{D^{*}}-m_{D})^2$, a bound state with mass 7189.7
MeV is found for $\Lambda=850$ MeV, However, this solution
disappears if one chooses $\mu^2=m^2_{ex}-(m_{B^{*}}-m_{B})^2$. Only
when $\Lambda$ is around 880 MeV, the bound state solutions can be
found for both $\mu^2$ choices. The difference of the static
properties for the two $\mu^2$ choices is relatively larger than
that of the above systems considered, this is because of the larger
difference between $m_{D^{*}}-m_{D}\simeq140$ MeV and
$m_{B^{*}}-m_{B}\simeq45$ MeV. We notice that the $D^{0}B^{*+}$
component has the largest probability in the states, since the
threshold of $D^{0}B^{*+}$ is lower than that of $D^{+}B^{*0}$,
$D^{*+}B^{0}$ and $D^{*0}B^{+}$. The wavefunction of the state with
mass about 7185.9 MeV and $\Lambda=900$ MeV is shown in Fig.
\ref{wavefunction of C=B=1}, it is obvious all the eight components
of the spatial wavefunction have the same sign, consequently this
state would be isospin singlet in the isospin symmetry limit.
Similar pattern of bound state solutions is predicted if the
coupling constants except $g_{\pi NN}$ are reduced by half. This
state is difficult to be produced, since both $c$ and $\bar{b}$ have
to be produced simultaneously. The direct production of this state
at hadron collider such as LHC and Tevatron is most promising, and
the indirect production via top quark decay is a possible
alternative. Once produced, it should be very stable, $DB\pi$ and
$DB\gamma$ are the main decay channels.
\subsection{Pseudoscalar-vector system with $C=-B=1$}
The effective interaction potentials are induced by one boson
exchange between two antiquarks, therefore both the $\pi$ exchange
and $\omega$ exchange contributions give opposite sign between the
$C=B=1$ system and the $C=-B=1$ system, nevertheless the overall
signs of $\eta$, $\sigma$ and $\rho$ exchange potentials remain. We
have eight coupled channels as well, $(D^{+}B^{*-})_S$,
$(D^{+}B^{*-})_D$, $(D^{0}\bar{B}^{*0})_S$, $(D^{0}\bar{B}^{*0})_D$,
$(D^{*+}B^{-})_S$, $(D^{*+}B^{-})_D$, $(D^{*0}\bar{B}^{0})_S$ and
$(D^{*0}\bar{B}^{0})_D$ are involved. The numerical results are
given in Table \ref{static properties of C=-B=1 system}. It is
remarkable that the $\mu^2$ and $\Lambda$ dependence of the bound
state solutions is similar to the $C=B=1$ case. With the same
$\mu^2$ and $\Lambda$ values, the predictions for the static
properties of the two systems are not drastically different from
each other. Concretely for $\Lambda=900$ MeV and
$\mu^2=m^2_{ex}-(m_{D^{*}}-m_{D})^2$, we find a bound state with
mass 7187.6 MeV for the $C=-B=1$ system, and the mass of $C=B=1$
bound state is 7185.9 MeV, the difference is about 1.7 MeV. The
corresponding wavefunction with $\Lambda=900$ MeV is plotted in Fig.
\ref{wavefunction of C=-B=1}, which can be roughly obtained by
reversing the overall sign of the third, fourth, seventh and eighth
components of the $C=B=1$ system wavefunction in Fig.
\ref{wavefunction of C=B=1}. To understand the similarity of the
predictions for the $C=B=1$ and $C=-B=1$ system, we turn to the one
$\pi$ exchange model, the effective potential comprises a spin-spin
potential proportional to
$(\bm{\sigma}_i\cdot\bm{\sigma}_j)(\bm{\tau}_i\cdot\bm{\tau}_j)$ and
a tensor potential proportional to
$S_{ij}(\hat{\mathbf{r}})(\bm{\tau}_i\cdot\bm{\tau}_j)$, where the
isospin matrix $\bm{\tau}_i$ and the spin matrix $\bm{\sigma}_i$
only act on the light quarks. In the basis of the eight channels
listed above, these two operators can be written as $8\times 8$
matrices
\begin{eqnarray}
\nonumber&&
(\bm{\sigma}_i\cdot\bm{\sigma}_j)(\bm{\tau}_i\cdot\bm{\tau}_j)\longrightarrow\left(\begin{array}{cccccccc}
0&0&0&0&-1&0&2&0\\
0&0&0&0&0&-1&0&2\\
0&0&0&0&2&0&-1&0\\
0&0&0&0&0&2&0&-1\\
-1&0&2&0&0&0&0&0\\
0&-1&0&2&0&0&0&0\\
2&0&-1&0&0&0&0&0\\
0&2&0&-1&0&0&0&0
\end{array}\right)\\
\label{21}&&
S_{ij}(\hat{\mathbf{r}})(\bm{\tau}_i\cdot\bm{\tau}_j)\longrightarrow\left(\begin{array}{cccccccc}
0&0&0&0&0&\sqrt{2}&0&-2\sqrt{2}\\
0&0&0&0&\sqrt{2}&-1&-2\sqrt{2}&2\\
0&0&0&0&0&-2\sqrt{2}&0&\sqrt{2}\\
0&0&0&0&-2\sqrt{2}&2&\sqrt{2}&-1\\
0&\sqrt{2}&0&-2\sqrt{2}&0&0&0&0\\
\sqrt{2}&-1&-2\sqrt{2}&2&0&0&0&0\\
0&-2\sqrt{2}&0&\sqrt{2}&0&0&0&0\\
-2\sqrt{2}&2&\sqrt{2}&-1&0&0&0&0
\end{array}\right)
\end{eqnarray}
For the $C=B=1$ pseudoscalar-vector system, the corresponding matrix
representations are obtained by replacing 2 and $2\sqrt{2}$ with -2
and $-2\sqrt{2}$ respectively in Eq.(\ref{21}). It is obvious both
operators contribute to only the off-diagonal $4\times4$ matrix
elements. As a result, the eigenvalues of the corresponding
Schr$\ddot{\rm o}$dinger equation for the $C=B=1$ and $C=-B=1$ cases
are exactly the same, if the small mass difference within the
isospin multiplets is neglected, and the eingen-wavefunction of one
system can be obtained from another by reversing the overall sign of
the third, fourth, seventh and eighth components. Therefore the
heavy bosons $\eta$, $\sigma$, $\rho$ and $\omega$ exchange
contributes to effective potential, and the pion exchange
contribution is still dominant. In short summary, even after
including shorter distance contributions from $\eta$, $\sigma$,
$\rho$ and $\omega$ exchange, the results obtained are qualitatively
the same as those in the one $\pi$ exchange model. The same
conclusion has been reached for all the system consider above.

\section{Conclusions and discussions}

Motivated by the nucleon-nucleon interaction, we have represented
the short range interaction by heavier mesons $\eta$, $\sigma$,
$\rho$ and $\omega$ exchange. The effective potentials between two
hadrons are obtained by summing the interactions between light
quarks or antiquarks via one boson exchange. The potential becomes
more complicated than that in the one pion exchange model, and there
are six additional terms which are proportional to $\mathbf{1}$,
$\bm{\tau}_i\cdot\bm{\tau}_j$, $\bm{\sigma}_i\cdot\bm{\sigma}_j$,
$S_{ij}(\hat{\mathbf{r}})$, $\mathbf{L}\cdot\mathbf{S}_{ij}$ and
$(\mathbf{L}\cdot\mathbf{S}_{ij})(\bm{\tau}_i\cdot\bm{\tau}_j)$
respectively.

We first apply the one boson exchange formalism to the deuteron,
then generalize to $D\bar{D}^{*}/\bar{D}D^{*}$,
$B\bar{B}^{*}/\bar{B}B^{*}$, $DD^{*}$, $BB^{*}$, PV systems with
$C=B=1$ and $C=-B=1$. S-D mixing effects has been taken into
account, and the uncertainties from the regularization parameter
$\Lambda$ and effective coupling constants are considered. We find
the conclusions reached are qualitatively the same as those in the
one pion exchange model. This implies that the long range $\pi$
exchange effects dominate the physics of a weakly bound hadronic
molecule, and we can safely use one pion exchange model to
qualitatively discuss the binding of molecule candidates. Since the
predictions for the binding energy and other static properties are
sensitive to the regularization parameter $\Lambda$ and the
effective couplings, we are not able to predict the binding energies
very precisely. If the potential is so strong that binding energy is
large enough, we would be quite confident that such bound state must
exist. However, the exact binding energy will depend on the details
of the regularization and the effective couplings involved. Our
results indicate that the $1^{++}$ $B\bar{B}^{*}/\bar{B}B^{*}$
molecule should exist, whereas $DD^{*}$ bound state doesn't exist.
For $\Lambda=808$ MeV (970 MeV), the binding energy, D wave
probability and other static properties of deuteron are produced,
meanwhile near threshold $1^{++}$ $D\bar{D}^{*}/\bar{D}D^{*}$
molecule is predicted. To identify this state with X(3872), the
mixing between this $D\bar{D}^{*}/\bar{D}D^{*}$ molecule and the
conventional charmonium state should be further considered to be
consistent with the recent experimental data on
$X(3872)\rightarrow\psi(2S)\gamma$ \cite{Fulsom:2008rn}. For the
$BB^{*}$ system, the PV systems with $C=B=1$ and $C=-B=1$, near
threshold molecular states may exist. Similar to the $1^{++}$
$D\bar{D}^{*}/\bar{D}D^{*}$ molecule, these states should be rather
stable, isospin is drastically broken, and the $I=0$ component is
dominant. Direct production of the above doubly heavy states at
Tevatron and LHC is the most promising way. We can search for the
$1^{++}$ $B\bar{B}^{*}/\bar{B}B^{*}$ molecule via
$p\bar{p}\rightarrow\pi^{+}\pi^{-}\Upsilon(1S)$ at Tevatron. The
$BB^{*}$ bound state mainly decays into $BB\gamma$ if it really
exists. The dominant decay channels of the heavy flavor PV bound
state with $C=B=1$ are $DB\pi$ and $DB\gamma$, and the possible
heavy flavor PV bound state with $C=-B=1$ mainly decays into
$D\bar{B}\pi$ and $D\bar{B}\gamma$.

In our model, the involved parameters include the effective
quark-boson couplings, the masses of the exchanged bosons and the
hadrons inside the molecule. Therefore this model is quite general,
it can be widely used to dynamically study the possible molecular
candidates. We will further apply the one boson exchange model to
baryon-antibaryon system etc, and compare the predictions with the
recent experimental observations \cite{progress}.

\begin{acknowledgments}
We acknowledge Prof. Dao-Neng Gao for stimulating discussions. This
work is supported by the China Postdoctoral Science foundation
(20070420735). Jia-Feng Liu is supported in part by the National
Natural Science Foundation of China under Grant No.10775124.
\end{acknowledgments}

\begin{appendix}
\section{The matrix elements of the spin relevant operators\label{appendix}}
For initial state consisting of two mesons $A$ and $B$, with
relative angular momentum $L$, total spin $S$ and total angular
momentum $J$, its wavefunction is written as
\begin{eqnarray}
\nonumber&&|(AB)LS,JM_J\rangle=\sum_{M_L,M_S} \langle
LM_L;SM_S|JM_J\rangle\; |LM_L\rangle|SM_S\rangle\\
\label{a1}&&=\sum_{S_{13},S_{24}}\hat{S}_A\hat{S}_B\hat{S}_{13}\hat{S}_{24}\left\{\begin{array}{ccc}
1/2&1/2&S_A\\
1/2&1/2&S_B\\
S_{13}&S_{24}&S
\end{array}\right\}|L(S_{13}S_{24})S,JM_J\rangle
\end{eqnarray}
where $\hat{S}=\sqrt{2S+1}$. For the convenience of calculating the
matrix elements of the spin-orbit operator
$\mathbf{L}\cdot\mathbf{S}_{24}$, we can recouple the state as
\begin{eqnarray}
\nonumber|(AB)LS,JM_J\rangle=&&\sum_{S_{13},S_{24},J_{LS}}(-1)^{L+S+J}\hat{S}_A\hat{S}_B\hat{S}_{13}\hat{S}_{24}\hat{S}\hat{J}_{LS}\left\{\begin{array}{ccc}
L&S_{24}&J_{LS}\\
S_{13}&J&S
\end{array}\right\}\left\{\begin{array}{ccc}
1/2&1/2&S_A\\
1/2&1/2&S_B\\
S_{13}&S_{24}&S
\end{array}\right\}\\
\label{a2}&&\times|(LS_{24})J_{LS}S_{13},JM_{J}\rangle
\end{eqnarray}
In the same way, we can recouple the the final state
$|(A'B')L'S',J'M'_J\rangle$ via the Wigner 6-j and 9-j coefficients.
In the following, we shall present the matrix elements of four light
quark operators involved in the work, which is helpful to
calculating the matrix representation of the effective interactions.
\begin{enumerate}
\item{The unit operator $\mathbf{1}$ }\\
Using Eq.(\ref{a1}), it is obvious that
\begin{eqnarray}
\nonumber&&\label{a3}\langle(A'B')L'S',J'M'_J|\mathbf{1}|(AB)LS,JM_J\rangle=\delta_{LL'}\delta_{SS'}\delta_{JJ'}\delta_{M_JM'_J}\sum_{S_{13},S_{24}}\hat{S}_A
\hat{S}'_{A}\hat{S}_{B}\hat{S}'_{B}\hat{S}^2_{13}\hat{S}^2_{24}\\
\label{a3}&&\times\left\{\begin{array}{ccc}
1/2&1/2&S_A\\
1/2&1/2&S_B\\
S_{13}&S_{24}&S
\end{array}\right\}
\left\{\begin{array}{ccc}
1/2&1/2&S'_A\\
1/2&1/2&S'_B\\
S_{13}&S_{24}&S
\end{array}\right\}=\delta_{LL'}\delta_{SS'}\delta_{S_AS'_{A}}\delta_{S_BS'_{B}}\delta_{JJ'}\delta_{M_JM'_J}
\end{eqnarray}
\item{The spin-spin operator $\bm{\sigma}_2\cdot\bm{\sigma}_4$}
\begin{eqnarray}
\nonumber&&\langle(A'B')L'S',J'M'_J|\bm{\sigma}_2\cdot\bm{\sigma}_4|(AB)LS,JM_J\rangle=\delta_{LL'}\delta_{SS'}\delta_{JJ'}\delta_{M_JM'_J}\sum_{S_{13},S_{24}}\hat{S}_A\hat{S}'_{A}
\hat{S}_B\hat{S}'_{B}\hat{S}^2_{13}\hat{S}^2_{24}\\
\label{a4}&&\times[2S_{24}(S_{24}+1)-3]\left\{\begin{array}{ccc}
1/2&1/2&S_A\\
1/2&1/2&S_B\\
S_{13}&S_{24}&S
\end{array}\right\}
\left\{\begin{array}{ccc}
1/2&1/2&S'_A\\
1/2&1/2&S'_B\\
S_{13}&S_{24}&S
\end{array}\right\}
\end{eqnarray}
where the spin operators $\bm{\sigma}_2$ and $\bm{\sigma}_4$ only
act on the light quarks and antiquarks

\item{The spin-orbit operator $\mathbf{L}\cdot\mathbf{S}_{24}$}
\begin{eqnarray}
\nonumber&&\langle(A'B')L'S',J'M'_J|\mathbf{L}\cdot\mathbf{S}_{24}|(AB)LS,JM_J\rangle=\delta_{LL'}\delta_{JJ'}\delta_{M_JM'_J}\sum_{S_{13},S_{24},J_{LS}}(-1)^{S+S'+2L+2J}\hat{S}_A
\hat{S}'_A\\
\nonumber&&\times\hat{S}_B\hat{S}'_B\hat{S}\hat{S}'\hat{S}^2_{13}\hat{S}^2_{24}\hat{J}^2_{LS}\frac{1}{2}[J_{LS}(J_{LS}+1)-L(L+1)-S_{24}(S_{24}+1)]\left\{\begin{array}{ccc}
L&S_{24}&J_{LS}\\
S_{13}&J&S
\end{array}\right\}
\\
\label{a5}&&\times\left\{\begin{array}{ccc}
L&S_{24}&J_{LS}\\
S_{13}&J&S'
\end{array}\right\}\left\{\begin{array}{ccc}
1/2&1/2&S_A\\
1/2&1/2&S_B\\
S_{13}&S_{24}&S
\end{array}\right\}
\left\{\begin{array}{ccc}
1/2&1/2&S'_A\\
1/2&1/2&S'_B\\
S_{13}&S_{24}&S
\end{array}\right\}
\end{eqnarray}
where $\mathbf{S}_{24}=\frac{1}{2}(\bm{\sigma}_2+\bm{\sigma}_4)$,
$\mathbf{L}$ is the relative spatial angular momentum. The matrix
elements of $\mathbf{L}\cdot\mathbf{S}_{24}$ can be calculated by
the Wigner-Echart theorem \cite{angular}, and the same result has
been obtained.
\item{The tensor operator $S_{24}(\hat{\mathbf{r}})\equiv3(\bm{\sigma}_2\cdot\hat{\bf r})(\bm{\sigma}_4\cdot\hat{\bf
r})-\bm{\sigma}_2\cdot\bm{\sigma}_4$}\\
It can be checked that the tensor operator $S_{24}(\hat{r})$ is
proportional to the scalar product of two rank-2 tensor operators
$Y_{2m}$ and $S^{(2)}_m$ with $m=0,\pm1,\pm2$, where $Y_{2m}$ is the
spherical harmonic function of degree 2, and the five components of
$S^{(2)}_m$ are
\begin{eqnarray}
\nonumber&&S^{(2)}_2=\frac{1}{2}S_{2+}S_{4+},~~~S^{(2)}_1=-\frac{1}{2}(S_{20}S_{4+}+S_{2+}S_{40}),~~~S^{(2)}_0=-\frac{\sqrt{6}}{12}(S_{2-}S_{4+}-4S_{20}S_{40}+S_{2+}S_{4-})\\
\label{a6}&&S^{(2)}_{-1}=\frac{1}{2}(S_{2-}S_{40}+S_{20}S_{4-}),~~~S^{(2)}_{-2}=\frac{1}{2}S_{2-}S_{4-}
\end{eqnarray}
Here $S_{2+}=\frac{1}{2}(\sigma_{2x}+i\sigma_{2y})$,
$S_{20}=\frac{1}{2}\sigma_{20}$ and
$S_{2-}=\frac{1}{2}(\sigma_{2x}-i\sigma_{2y})$. The same convention
applies to $S_{4,\pm}$ and $S_{40}$, the spin operators
$\bm{\sigma}_2$ and $\bm{\sigma}_4$ only act on the light quark and
antiquarks. Using the Wigner-Echart theorem, the matrix element of
this tensor operator can be obtained, although it is somewhat
lengthy.
\begin{eqnarray}
\nonumber&&\langle(A'B')L'S',J'M'_J|S_{24}(\hat{\mathbf{r}})|(AB)LS,JM_J\rangle=\delta_{JJ'}\delta_{M_JM'_J}\frac{2}{3}\sqrt{30}\sum_{S_{13},S_{24}}\delta_{S_{24},1}(-1)^{J+L+L'+2S'+S_{13}+S_{24}}\\
\nonumber&&\times\hat{S}_A\hat{S}'_A\hat{S}_B\hat{S}'_B\hat{S}\hat{S}'\hat{L}\hat{L}'\hat{S}^2_{13}\hat{S}^4_{24}\left\{\begin{array}{ccc}
L'&S'&J\\
S&L&2
\end{array}\right\}\left\{\begin{array}{ccc}
S_{24}&S'&S_{13}\\
S&S_{24}&2
\end{array}\right\}\left(\begin{array}{ccc}
L'&2&L\\
0&0&0
\end{array}\right)\left\{\begin{array}{ccc}
1/2&1/2&S_A\\
1/2&1/2&S_B\\
S_{13}&S_{24}&S
\end{array}\right\}\\
\label{a7}&&\times\left\{\begin{array}{ccc}
1/2&1/2&S'_A\\
1/2&1/2&S'_B\\
S_{13}&S_{24}&S'
\end{array}\right\}
\end{eqnarray}
The above expression is apparently different from the results in
Ref. \cite{Thomas:2008ja}, However, the numerical results of all the
matrix elements are the same.
\end{enumerate}

\end{appendix}

\newpage

\begin{figure}[hptb]
\begin{center}
\begin{tabular}{cc}
\includegraphics[scale=.645]{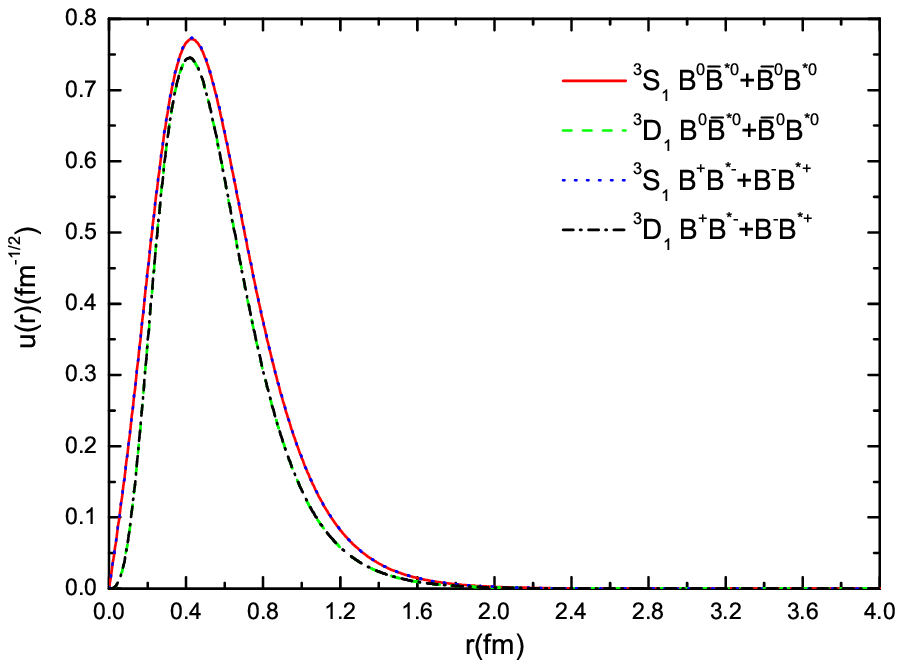}&\includegraphics[scale=.645]{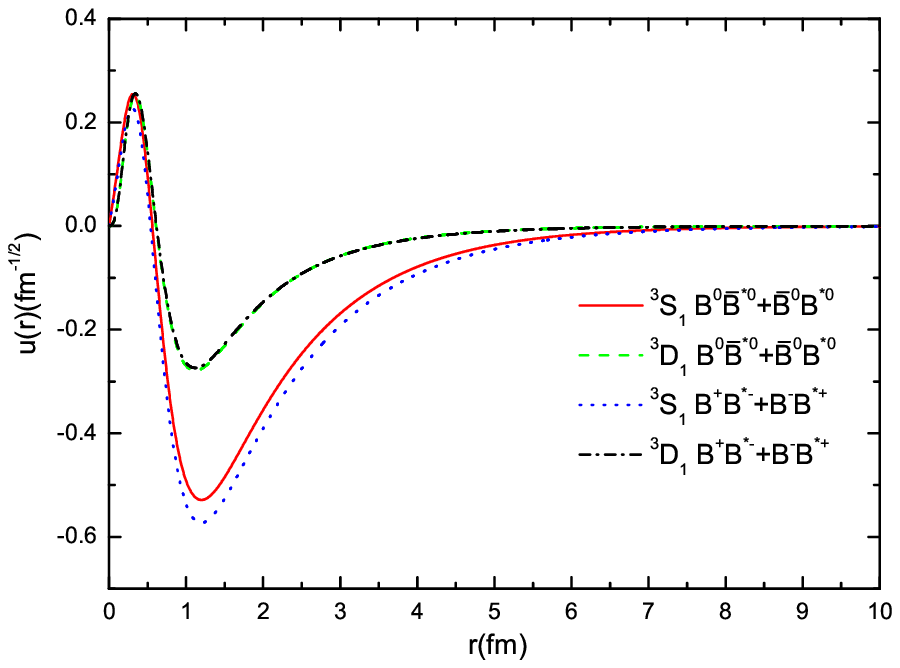}\\
(a)&(b)
\end{tabular}
\caption{\label{wavefunction of bottom analog} The spatial
wavefunctions of $1^{++}$ $B\bar{B}^{*}/\bar{B}B^{*}$ molecule with
$\Lambda=1000$ MeV. There are two bound states with mass 10457.6 MeV
and 10600.4 MeV respectively, Fig. \ref{wavefunction of bottom
analog}a is for the first state, and the Fig. \ref{wavefunction of
bottom analog}b for the second.}
\end{center}
\end{figure}

\begin{figure}[hptb]
\includegraphics[scale=.645]{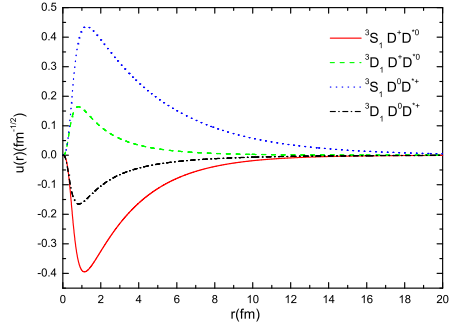}
\caption{\label{wavefunction of C=2} The wavefunction for the
$DD^{*}$ system assuming $\Lambda=1600$ MeV and
$\mu^2=\mu^2_{cc1}\equiv m^2_{\pi^{0}}-(m_{D^{*+}}-m_{D^{+}})^2$,
its mass approximately is 3873.9 MeV.}
\end{figure}

\begin{figure}[hptb]
\includegraphics[scale=.645]{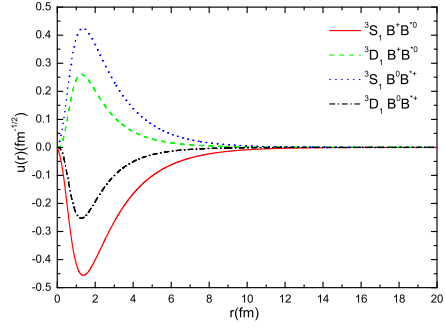}
\caption{\label{wavefunction of B=2} The spatial wavefunction for
the $BB^{*}$ system assuming $\Lambda=900$ MeV and
$\mu^2=\mu^2_{bb1}\equiv m^2_{\pi^{0}}-(m_{B^{*}}-m_{B^{+}})^2$, its
mass is 10602.3 MeV.}
\end{figure}

\begin{figure}[hptb]
\includegraphics[scale=.645]{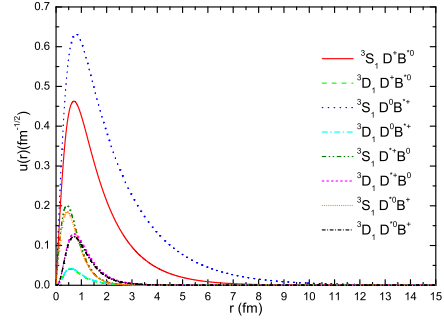}
\caption{\label{wavefunction of C=B=1} The wavefunction of the
$C=B=1$ pseudoscalar-vector system with $\Lambda=900$ MeV and
$\mu^2=m^2_{ex}-(m_{D^{*}}-m_{D})^2$, the mass of this state is
about 7185.9 MeV.}
\end{figure}

\begin{figure}[hptb]
\includegraphics[scale=.645]{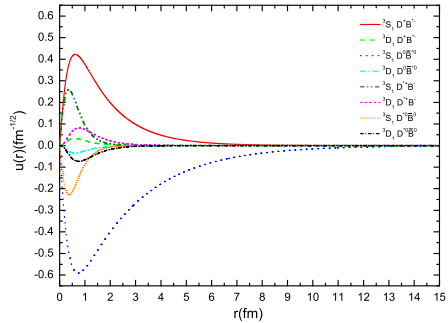}
\caption{\label{wavefunction of C=-B=1} The wavefunction of the
$C=-B=1$ pseudoscalar-vector system with $\Lambda=900$ MeV and
$\mu^2=m^2_{ex}-(m_{D^{*}}-m_{D})^2$, its mass is about 7187.6 MeV.}
\end{figure}

\newpage

\begin{center}
\begin{table}
\begin{tabular}{|c|cccc|}\hline\hline

\multicolumn{5}{|c|}{$1^{++}$
$B\bar{B}^{*}/\bar{B}B^{*}$}\\\hline\hline

$\mu^2$ & $\Lambda({\rm MeV})$&$~~~{\rm M}(\rm MeV)$&$~~~{\rm
r}_{\rm
rms}({\rm fm})$&$~~~{\rm P^{00}_S:P^{00}_D:P^{+-}_S:P^{+-}_D}(\%)$\\
\hline\hline

 & 808 &  10565.3    & 0.60 &  47.70:2.05:48.20:2.05   \\

$\mu^2_{b\bar{b}1}$ & 900& 10543.5&  0.59    &  44.11:5.69:44.50:5.70     \\

 & 1000& 10457.6    & 0.52    &  27.52:22.41:27.64:22.44  \\
  &   &10600.4    &   1.73&  37.10:9.43:44.26:9.22  \\ \hline

 &808 & 10565.3    &  0.60 & 47.70:2.05:48.20:2.05  \\

$\mu^2_{b\bar{b}2}$ &900& 10543.5  &  0.59  &   44.11:5.69:44.49:5.70    \\

 & 1000& 10457.6   &   0.52   &  27.52:22.41:27.64:22.44  \\
 &  &10600.4     &   1.72    &    37.10:9.43:44.25:9.22 \\ \hline\hline

\multicolumn{5}{|c|}{all coupling except $g_{\pi NN}$ are reduced by
half}\\\hline\hline

$\mu^2$& $\Lambda({\rm MeV})$&$~~~{\rm M}(\rm MeV)$&$~~~{\rm r}_{\rm
rms}({\rm fm})$&$~~~{\rm
P^{00}_S:P^{00}_D:P^{+-}_S:P^{+-}_D}(\%)$\\\hline

 & 970 & 10544.8 &  0.55 & 45.24:4.60:45.55:4.61  \\

$\mu^2_{b\bar{b}1}$ &1100 &10503.9 &  0.51 & 40.12:9.78:40.30:9.80  \\


 & 1200  & 10443.1  & 0.46     &  33.66:16.27:33.77:16.29   \\
 &     & 10601.9  & 1.91 & 38.82:8.03:45.26:7.89 \\\hline

 & 970 & 10544.8  &   0.55  &   45.24:4.60:45.55:4.61  \\

 $\mu^2_{b\bar{b}2}$& 1100 & 10503.9  &   0.51   &   40.12:9.78:40.30:9.80    \\


 &1200  &  10443.1  & 0.46    & 33.66:16.27:33.77:16.29   \\
 &     &10601.9 &   1.91& 38.83:8.04:45.25:7.89
  \\\hline\hline

\end{tabular}
\caption{\label{bottom analog of x3872}The predictions about the
mass, the root of mean square radius(rms) and the probabilities of
the different components for the $1^{++}$
$B\bar{B}^{*}/\bar{B}B^{*}$ system with
$\mu^2_{b\bar{b}1}=m^2_{\pi^{\pm}}-(m_{B^{*}}-m_{B^{0}})^2$ and
$\mu^2_{b\bar{b}2}=m^2_{\pi^{\pm}}-(m_{B^{*}}-m_{B^{+}})^2$.}
\end{table}
\end{center}

\begin{center}
\begin{table}[hptb]
\begin{tabular}{|c|cccc|}\hline\hline

\multicolumn{5}{|c|}{${DD^{*}}$ system with $C=2$}\\\hline\hline

$\mu^2$ & $\Lambda({\rm MeV})$&$~~~{\rm M}(\rm MeV)$&$~~~{\rm
r}_{\rm rms}({\rm fm})$&$~~~{\rm
P^{+0}_S:P^{+0}_D:P^{0+}_S:P^{0+}_D}(\%)$\\\hline

 &1600&  3873.9     &   3.15     &    34.61:4.03:56.82:4.54    \\

$\mu^2_{cc1}$ &1700&   3865.1   & 1.30     &    20.29:28.37:22.63:28.71        \\

 &1800& 3770.9     &  0.58      &    0.19:49.78:0.19:49.84    \\
&  & 3872.8       &  2.48    &    41.47:1.38:55.54:1.63 \\
\hline

 & 1600&  3873.9         & 3.16           &  34.54:4.04:56.87:4.55     \\

$\mu^2_{cc2}$ & 1700&  3865.1       & 1.30        &   20.31:28.35:22.65:28.70         \\

 & 1800&   3771.0     &   0.58      &      0.19:49.78:0.19:49.84       \\
  &  &   3872.8    &   2.49     &  41.41:1.38:55.58:1.63\\
\hline\hline

\multicolumn{5}{|c|}{all couplings are reduced by half except
$g_{\pi NN}$}\\\hline\hline

$\mu^2$ &$\Lambda({\rm MeV})$&$~~~{\rm M}(\rm MeV)$&$~~~{\rm r}_{\rm
rms}({\rm fm})$&$~~~{\rm
P^{+0}_S:P^{+0}_D:P^{0+}_S:P^{0+}_D}(\%)$\\\hline

 & 1900&   3873.1  & 2.53     &  36.05:5.90:51.55:6.50    \\

$\mu^2_{cc1}$ &2000&   3870.1  & 1.82  &   38.09:8.08:45.32:8.52    \\

 & 2100&  3865.3   & 1.44  &   37.67:10.21:41.57:10.55  \\

 & 2200&    3858.4  &  1.19   &    36.31:12.42:38.60:12.68  \\
\hline

 &1900&  3873.1    &  2.54      &  36.01:5.91:51.58:6.51  \\

$\mu^2_{cc2}$ &2000&  3870.1    & 1.82   &    38.07:8.08:45.32:8.53  \\

 &2100&  3865.4  & 1.44    &   37.66:10.22:41.57:10.55   \\

 & 2200&   3858.4   &  1.19   &  36.30:12.42:38.59:12.68
\\ \hline\hline

\end{tabular}
\caption{\label{static properties of C=2 system}The predictions
about the mass, rms and the probabilities of the different
components for the ${DD^{*}}$ system, and ${\rm P^{+0}_S}$ denotes
the probability of S wave ${D^{+}D^{*0}}$ in the state, and the
meaning of ${\rm P^{+0}_D}$, ${\rm P^{0+}_S}$ and ${\rm P^{0+}_D}$
is similar. Here
$\mu^2_{cc1}=m^2_{\pi^{0}}-(m_{D^{*+}}-m_{D^{+}})^2$ and
$\mu^2_{cc2}=m^2_{\pi^{0}}-(m_{D^{*0}}-m_{D^{0}})^2$ }
\end{table}
\end{center}

\begin{center}
\begin{table}[hptb]
\begin{tabular}{|c|cccc|}\hline\hline

\multicolumn{5}{|c|}{${BB^{*}}$ system with $B=2$}\\\hline\hline

$\mu^2$ & $\Lambda({\rm MeV})$&$~~~{\rm M}(\rm MeV)$&$~~~{\rm
r}_{\rm rms}({\rm fm})$&$~~~{\rm
P^{+0}_S:P^{+0}_D:P^{0+}_S:P^{0+}_D}(\%)$\\\hline

 &808&  10603.9   &    4.09      &   59.37:6.74:28.49:5.40         \\

$\mu^2_{bb1}$ &900& 10602.3 &      2.23        &  43.52:10.84:35.63:10.02\\

 &1000& 10598.8 &     1.61         &  37.22:14.37:34.48:13.93 \\

 & 1100&   10592.2    &     1.27   & 31.33:19.34:30.24:19.09
\\\hline

 & 808&  10603.9       &    4.09  &     59.38:6.74:28.48:5.40   \\

$\mu^2_{bb2}$ & 900&10602.3      &        2.23  &  43.52:10.84:35.63:10.01 \\

 &1000& 10598.8 & 1.61    &  37.22:14.37:34.48:13.93  \\

 & 1100&  10592.2     & 1.27    &     31.33:19.34:30.24:19.09        \\
\hline\hline

\multicolumn{5}{|c|}{all couplings except $g_{\pi NN}$ are reduced
by half}\\\hline\hline

$\mu^2$ &$\Lambda({\rm MeV})$&$~~~{\rm M}(\rm MeV)$&$~~~{\rm r}_{\rm
rms}({\rm fm})$&$~~~{\rm
P^{+0}_S:P^{+0}_D:P^{0+}_S:P^{0+}_D}(\%)$\\\hline

 & 970&   10601.5        &   1.99      &    39.93:13.08:34.67:12.31      \\

$\mu^2_{bb1}$ &1000  & 10600.7     &   1.82     &  38.34:14.02:34.30:13.34\\

 &1100&  10596.6      &   1.44     &    34.37:16.80:32.47:16.36      \\

 &1200& 10590.3    &   1.19        &     31.33:19.33:30.32:19.03           \\
\hline

 & 970&  10601.5       & 1.99      &    39.94:13.08:34.67:12.31        \\

$\mu^2_{bb2}$ &1000&  10600.7       &    1.82     &    38.34:14.02:34.30:13.34          \\

 &1100&   10596.6      &  1.44        &   34.37:16.80:32.47:16.36           \\

 &1200&    10590.3         & 1.19          &   31.33:19.33:30.32:19.03          \\
\hline\hline

\end{tabular}
\caption{\label{static properties of B=2 system}The predictions
about the mass, rms and the probabilities of the different
components for the ${\rm BB^{*}}$ system with B=2. ${\rm P^{+0}_S}$
represents the probability of S-wave ${\rm B^{+}B^{*0}}$. Here
$\mu^2_{bb1}=m^2_{\pi^{0}}-(m_{B^{*}}-m_{B^{+}})^2$ and
$\mu^2_{bb2}=m^2_{\pi^{0}}-(m_{B^{*}}-m_{B^{0}})^2$.}
\end{table}
\end{center}

\begin{center}
\begin{table}[hptb]
\begin{tabular}{|c|cccc|}\hline\hline

\multicolumn{5}{|c|}{\scriptsize The pseudoscalar-vector system with
$C=B=1$}\\\hline\hline

{\scriptsize $\mu^2$} & {\scriptsize  $ \Lambda({\rm
MeV})$}&{\scriptsize ${\rm M}(\rm MeV)$}&{\scriptsize ${\rm r}_{\rm
rms}({\rm fm})$}&{\scriptsize $~~{\rm
P^{+0}_S(DB^{*}):P^{+0}_D(DB^{*}):P^{0+}_S(DB^{*}):P^{0+}_D(DB^{*}):P^{+0}_S(D^{*}B):P^{+0}_D(D^{*}B):P^{0+}_S(D^{*}B):P^{0+}_D(D^{*}B)}(\%)$}\\\hline

 &850 & 7189.7   &  5.54  & 8.10:0.02:89.69:0.02:0.87:0.36:0.67:0.26   \\

 {\scriptsize $\mu^2_{\bar{b}c1}$}&880 & 7187.9 & 2.05   & 21.88:0.07:72.40:0.08:2.09:0.96:1.73:0.80\\

 &900 & 7185.9  & 1.58  & 27.34:0.11:65.19:0.12:2.54:1.35:2.16:1.19   \\

 & 1000 & 7157.2   &  0.86  & 36.55:0.01:46.27:0.01:1.39:7.32:1.21:7.26   \\ \hline

 & 850 & no bounded  & --- & --- \\

{\scriptsize $\mu^2_{\bar{b}c2}$} & 880 &  7189.5   &  4.11     &  10.70:0.02:86.98:0.02:0.79:0.50:0.61:0.39   \\

 &900 &  7188.1  &    2.18    &  20.59:0.04:75.09:0.04:1.34:0.98:1.10:0.83    \\

 & 1000 & 7161.1    &   0.88   &   36.86:0.22:47.18:0.23:0.48:7.33:0.41:7.31  \\
\hline\hline

\multicolumn{5}{|c|}{\scriptsize all couplings are reduced by half
except $g_{\pi NN}$}\\\hline\hline

{\scriptsize $\mu^2$}  &{\scriptsize$\Lambda({\rm
MeV})$}&{\scriptsize${\rm M}(\rm MeV)$}&{\scriptsize${\rm r}_{\rm
rms}({\rm fm})$}&{\scriptsize $~~{\rm
P^{+0}_S(DB^{*}):P^{+0}_D(DB^{*}):P^{0+}_S(DB^{*}):P^{0+}_D(DB^{*}):
P^{+0}_S(D^{*}B):P^{+0}_D(D^{*}B):P^{0+}_S(D^{*}B):P^{0+}_D(D^{*}B)}(\%)$}\\\hline

 & 970 & 7189.2     &   3.13          &   16.76:0.07:78.38:0.08:1.75:0.85:1.45:0.67   \\

{\scriptsize $\mu^2_{\bar{b}c1}$} & 1000 & 7187.6   &   1.88       &   25.51:0.13:66.52:0.15:2.78:1.35:2.42:1.15\\

 & 1100 & 7177.1   &   1.03      &  34.50:0.42:49.32:0.45:4.94:2.96:4.62:2.81\\

 & 1200 &  7156.3    &   0.78 &   34.62:0.72:41.51:0.75:5.85:5.51:5.64:5.42
\\ \hline

 & 970 & no bounded   &    ---          &  --- \\

{\scriptsize $\mu^2_{\bar{b}c2}$} &1000 &  7189.6    &  4.90  & 11.19:0.03:86.02:0.03:0.90:0.62:0.74:0.49\\

 & 1100 &  7181.9    &  1.20   &  34.04:0.20:54.01:0.21:3.26:2.67:3.08:2.53  \\

 & 1200 &  7163.0   &  0.83     &  36.21:0.33:43.97:0.34:4.08:5.57:4.00:5.51
\\ \hline\hline

\end{tabular}
\caption{\label{static properties of C=B=1 system}The predictions
for the static properties of the PV system with ${C=B=1}$, where
${\rm P^{+0}_S(DB^{*})}$ denotes the probability of S-wave ${
D^{+}B^{*0}}$, and ${\rm P^{+0}_S(D^{*}B)}$ denotes the probability
of S-wave ${D^{*+}B^{0}}$. Here
$\mu^2_{\bar{b}c1}=m^2_{ex}-(m_{D^{*}}-m_{D})^2$ and
$\mu^2_{\bar{b}c2}=m^2_{ex}-(m_{B^{*}}-m_{B})^2$ with $m_{ex}$ the
exchanged boson mass.}
\end{table}
\end{center}

\begin{center}
\begin{table}[hptb]
\begin{tabular}{|c|cccc|}\hline\hline

\multicolumn{5}{|c|}{\scriptsize The pseudoscalar-vector system with
$C=-B=1$}\\ \hline\hline

{\scriptsize $\mu^2$ }& {\scriptsize $\Lambda({\rm
MeV})$}&{\scriptsize ${\rm M}(\rm MeV)$}&{\scriptsize ${\rm r}_{\rm
rms}({\rm fm})$}&{\scriptsize $~~~{\rm
P^{+-}_S(DB^{*}):P^{+-}_D(DB^{*}):P^{00}_S(DB^{*}):P^{00}_D(DB^{*}):P^{+-}_S(D^{*}B):P^{+-}_D(D^{*}B):P^{00}_S(D^{*}B):P^{00}_D(D^{*}B)}(\%)$}\\\hline

 & 880 &  7189.2      &    3.11       &  15.16:0.04:80.15:0.05:2.21:0.45:1.63:0.32          \\

{\scriptsize $\mu^2_{bc1}$} & 900 &  7187.6    &    1.85   &   23.82:0.08:68.07:0.09:3.83:0.61:3.02:0.48      \\

  & 1000 & 7169.0       &    0.79      &    30.56:0.28:46.30:0.30:11.86:0.74:9.27:0.69       \\

 & 1050 &   7148.4        &   0.51     &   15.57:0.02:53.99:0.04:22.84:0.10:7.41:0.03         \\
 &    &   7154.3       &  0.63      &   53.57:0.36:16.62:0.34:5.65:0.59:22.22:0.66            \\ \hline

 & 880 &     no bounded        &    ---            &     ---        \\

{\scriptsize $\mu^2_{bc2}$} & 900 &   7189.7         &   5.17   &   9.30:0.02:88.09:0.02:1.25:0.27:0.87:0.19                   \\

 & 1000&    7176.2      & 0.91      &  30.77:0.20:50.54:0.21:9.62:0.70:7.32:0.65                    \\

 & 1050 &  7154.6    &    0.52   &    25.22:0.00:45.64:0.01:17.96:0.04:11.14:0.00                         \\
  &   &   7163.3     &   0.70     &   46.16:0.31:27.72:0.30:7.95:0.64:16.25:0.68       \\
\hline\hline

\multicolumn{5}{|c|}{\scriptsize all couplings except $g_{\pi NN}$
are reduced half}\\ \hline\hline

{\scriptsize $\mu^2$} &{\scriptsize $\Lambda({\rm
MeV})$}&{\scriptsize ${\rm M}(\rm MeV)$}&{\scriptsize ${\rm r}_{\rm
rms}({\rm fm})$}&{\scriptsize $~~~{\rm
P^{+-}_S(DB^{*}):P^{+-}_D(DB^{*}):P^{00}_S(DB^{*}):P^{00}_D(DB^{*}):P^{+-}_S(D^{*}B):P^{+-}_D(D^{*}B):P^{00}_S(D^{*}B):P^{00}_D(D^{*}B)}(\%)$}\\\hline

 & 970 &    7189.4   &  3.57     &   15.68:0.06:78.98:0.07:2.29:0.58:1.90:0.44         \\

{\scriptsize $\mu^2_{bc1}$} &1020&   7185.6      &   1.38  &  30.23:0.20:56.36:0.21:5.82:1.03:5.25:0.90         \\

 & 1100 &  7173.2    &   0.82   &    33.18:0.42:43.08:0.45:10.53:1.24:9.93:1.17  \\

 & 1200 &    7147.4   &   0.59          &   30.96:0.68:35.69:0.70:15.05:1.28:14.40:1.25       \\
\hline

 & 970 &     no  bounded        &       ---       &      ---       \\

{\scriptsize $\mu^{2}_{bc2}$} & 1020 &   7189.2    &   3.03      &   18.64:0.07:75.25:0.07:2.60:0.63:2.24:0.51      \\

 & 1100 &   7180.5   &  1.00       & 33.61:0.29:47.78:0.30:8.09:1.16:7.68:1.09     \\

 & 1200 &   7158.2       &   0.64      &  32.34:0.58:37.41:0.60:13.43:1.28:13.09:1.26       \\
\hline\hline

\end{tabular}
\caption{\label{static properties of C=-B=1 system} The predictions
about the static properties of the PV system with ${C=-B=1}$, where
${\rm P^{+-}_S(DB^{*})}$ denotes the probability of S wave ${
D^{+}B^{*-}}$, and ${\rm P^{+-}_S(D^{*}B)}$ denotes the probability
of S wave ${ D^{*+}B^{-}}$. Here
$\mu^2_{bc1}=m^2_{ex}-(m_{D^{*}}-m_{D})^2$ and
$\mu^2_{bc2}=m^2_{ex}-(m_{B^{*}}-m_{B})^2$ with $m_{ex}$ the
exchanged boson mass.}
\end{table}
\end{center}

\end{document}